\documentclass[11pt,a4paper]{article}

\usepackage{jheppub}
\makeatletter
\gdef\@fpheader{}
\makeatother

\usepackage[english]{babel}
\usepackage[babel]{csquotes}
\usepackage{mathtools}
\usepackage{float}
\usepackage{braket}
\usepackage{booktabs}
\usepackage{nicematrix}
\usepackage{subcaption}
\usepackage{enumitem}
\usepackage[font=small,format=hang,labelfont={sf,bf},figurewithin=section,tablewithin=section]{caption}

\definecolor{darkblue}{rgb}{0.1,0.1,0.7}
\hypersetup{colorlinks,
           linkcolor={darkblue},
           citecolor={darkblue},
           urlcolor={darkblue}
}

\newcommand{\eps}{{\varepsilon}}

\renewcommand{\geq}{\geqslant}
\renewcommand{\leq}{\leqslant}
\newcommand{\ped}[1]{\textormath{\textsubscript{#1}}{_{\mathrm{#1}}}}
\newcommand{\ap}[1]{\textormath{\textsuperscript{#1}}{^{\mathrm{#1}}}}
\newcommand{\id}{\text{\usefont{U}{bbold}{m}{n}1}}
\newcommand{\tr}{\mathrm{tr}}

\newcommand{\lsp}{\hspace{1pt}}

\let\originalleft\left
\let\originalright\right
\renewcommand{\left}{\mathopen{}\mathclose\bgroup\originalleft}
\renewcommand{\right}{\aftergroup\egroup\originalright}

\makeatletter
\renewcommand*\env@matrix[1][*\c@MaxMatrixCols c]{
    \hskip -\arraycolsep
    \let\@ifnextchar\new@ifnextchar
    \array{#1}
}
\makeatother
\newcommand{\cg}[6]{
    \begin{Bmatrix}[cc|c]
      #1 & #3 & #5 \\
      #2 & #4 & #6
    \end{Bmatrix}
}

\newcommand{\defaultplotscale}{0.87}

\title{Exploring Replica-Potts CFTs in Two Dimensions}

\author{Stefanos R.\ Kousvos$^{a}$}
\author{Alessandro Piazza,$^{a,b,c,d}$}
\author{Alessandro Vichi$^{a}$}
\affiliation{
    $^a$ Department of Physics, University of Pisa and INFN, section of Pisa. Largo Pontecorvo 3, I-56127 Pisa, Italy\\
    $^b$ SISSA, Via Bonomea 265, I-34136 Trieste, Italy \\
    $^c$ INFN, Sezione di Trieste, Via Valerio 2, I-34127 Trieste, Italy \\
    $^d$ Scuola Normale Superiore, Piazza dei Cavalieri 7, I-56126 Pisa, Italy\\
}

\emailAdd{stefanos.kousvos@df.unipi.it} \emailAdd{alessandro.piazza@sissa.it} \emailAdd{alessandro.vichi@df.unipi.it}

\abstract{We initiate a numerical conformal bootstrap study of CFTs with $S_n \ltimes (S_Q)^n$ global symmetry. These include CFTs that can be obtained as coupled replicas of two-dimensional critical Potts models. Particular attention is paid to the special case $S_3 \ltimes (S_3)^3$, which governs the critical behaviour of three coupled critical 3-state Potts models, a multi-scalar realisation of a (potentially) non-integrable CFT in two dimensions. The model has been studied in earlier works using perturbation theory, transfer matrices, and Monte Carlo simulations. This work represents an independent non-perturbative analysis. Our results are in agreement with previous determinations: we obtain an allowed peninsula within parameter space for the scaling dimensions of the three lowest-lying operators in the theory, which contains the earlier predictions for these scaling dimensions. Additionally, we derive numerous bounds on admissible scaling dimensions in the theory, which are compatible with earlier results. Our work sets the necessary groundwork for a future precision study of these theories in the conformal bootstrap.
}

\begin{document}

\maketitle

\section{Introduction}
Following the resurgence of the conformal bootstrap program with~\citep{Rattazzi:2008pe}, a considerable amount of effort has been devoted to studying higher-dimensional CFTs ($d>2$). For a thorough account, see~\citep{Poland:2018epd}, and for a more recent account, see~\citep{Rychkov:2023wsd}. The flagship results of this program include particularly precise critical exponents for the Ising~\citep{Kos:2016ysd}, $O(2)$~\citep{Chester:2019ifh}, $O(3)$~\citep{Chester:2020iyt}, GNY~\citep{Erramilli:2022kgp} and super-Ising~\citep{Atanasov:2022bpi} universality classes. These theories correspond to three-dimensional CFTs which are not exactly solvable/solved, but are accessible via perturbative techniques such as the $\varepsilon$-expansion and non-perturbative techniques such as lattice simulations (see~\citep{Pelissetto:2000ek} for a review) and, more recently, also the fuzzy sphere regularisation technique~\citep{Zhu:2022gjc}. While the numerical conformal bootstrap method can be applied in $d=2$ using global conformal blocks, and the technology required to do this is indeed available, there are no analogues of the above-mentioned studies for two-dimensional theories. This is despite older literature~\citep{Dotsenko:1998gyp}, as well as more recent literature~\citep{Antunes:2022vtb}, suggesting the existence of families of interesting non-integrable CFTs. For a somewhat recent overview of $2d$ CFT see \citep{Yin:2017yyn}, for an alternative approach to the $4-$pt function bootstrap see \citep{Hellerman:2009bu}.

In this work, we take the first steps to remedy this by initiating a study of multi-scalar theories with $S_3 \ltimes (S_3)^3$ global symmetry, and more generally $S_n \ltimes (S_Q)^n$.\footnote{For the study of theories with $S_3$ global symmetry in $d=2$ and above, which corresponds to a single $3$-state Potts model, versus replicas which we will study in this work, see~\citep{Rong:2017cow,Chester:2022hzt}.} The universality class of these CFTs is supposed to be realized by coupled Potts models~\citep{Dotsenko:1998gyp}. It is believed to be among those $2d$ CFTs with the simplest spectrum, whose exact solution is not known. More pedantically, within the classification of $2d$ CFTs with central charge $c > 1$, this model is expected to be among the few examples of unitary, compact, irrational,\footnote{That is, with an infinite number of Virasoro primaries and without an extended algebra organizing them into a finite number of representations.} non-supersymmetric, and non-multicritical CFTs. The universality class can be understood as the IR fixed point of an ``energy-energy'' deformation around copies of critical Potts models
\begin{equation}
S = \sum_{a=1}^n S\ped{Potts}^{a} + g\int \mathrm{d}^d x \sum_{a \neq b} \epsilon^{a} \epsilon^{b} \, ,
\label{lagrangian}
\end{equation}
where $S\ped{Potts}^{a}$ is the action of the $a$-th replica of a $Q$-state Potts model, and the operators $\epsilon^{a}$ are the energy operators, i.e.\ the operators that would multiply the mass term.\footnote{This would for example be the operator $\phi^2$ in a perturbative treatment around the free theory. More generally, it's the lowest-dimensional non-trivial singlet of the single critical $Q$-state Potts model.} Finally, $g$ is a coupling which is set to zero at the decoupled fixed point. This deformation possesses controlled perturbative fixed points in the $Q=2+\varepsilon$ limit~\citep{Ludwig:1987rk,Ludwig:1989rj,Dotsenko:1994im,Dotsenko:1994sy}, where $\varepsilon$ is taken to be infinitesimal, and also in the large-$n$ limit (large number of replicas/copies). In the case of the large-$n$ limit, the computations have not been carried out explicitly for $S_n \ltimes (S_Q)^n$ in $d = 2$, however, they have been carried out to leading order for $S_n \ltimes (\mathbb{Z}_2)^n$ in $d = 3$~\citep{Binder:2021vep}. 
Notice also that this theory cannot be reached by a naive $d=d_c-\eps$ dimensional expansion, such as the $d=4-\eps$ expansion for $\phi^4$ theories, and the $d=6-\eps$ expansion for $\phi^3$ theories. This is due to the theory, morally speaking, being a $\phi^3 + \phi^4$ theory, the $\phi^3$ coming from each replica and the $\phi^4$ coming from the term that couples the replicas. This is in contrast, for example, with the critical $3$-state Potts model, which can in principle be treated with the $d=6-\eps$ expansion, albeit with a very long flow in terms of $\eps$, needed to reach $d=2$. The theory may however be amenable to a fixed dimension expansion in the spirit of~\citep{Serone:2018gjo}. 

Using a lattice Hamiltonian of the schematic type~\citep{Dotsenko:1998gyp}
\begin{equation}
    H = \sum_{\mathrm{n.n.}} \left[ a(\delta_{\sigma, \sigma^\prime}+\delta_{\tau, \tau^\prime}+\delta_{\eta, \eta^\prime})+b(\delta_{\sigma, \sigma^\prime}\delta_{\tau, \tau^\prime}+\delta_{\tau, \tau^\prime}\delta_{\eta, \eta^\prime}+\delta_{\eta, \eta^\prime}\delta_{\sigma, \sigma^\prime})\right] \, ,
\end{equation}
the theory can also be studied non-perturbatively. The sum runs over lattice sites and nearest neighbours (denoted with a prime) while $\sigma$, $\tau$ and $\eta$ represent spin configurations in each of the $n=3$ Potts replicas. For $Q = 3$, the spins in each replica can be thought of vectors pointing to any of the three vertices of an equilateral triangle. The term multiplying the coupling $a$ is just the sum of three decoupled Potts Hamiltonians, whereas $b$ multiplies the ``energy-energy'' interaction. The non-perturbative methods used in~\citep{Dotsenko:1998gyp}, such as transfer matrices and Monte Carlo simulations, also seem to indicate the existence of a fixed point. We will compare to these results in much of the present work, hence we postpone their discussion. 

It is currently believed that unitary non-trivial fixed points exist for $2<Q \leq 4$ and $n>2$. The pedestrian understanding for the constraint $2<Q \leq 4$ is that it guarantees the decoupled theories one perturbs of off are unitary. Given that the $\beta$ function is proportional to $\varepsilon g + A g^2+O(g^3)$ with $A$ a number, we do not get a controlled fixed point when $\varepsilon=0$ or equivalently when $Q=2$. As far as the limitation $n>2$ is concerned, we refer to the discussion in~\citep{Dotsenko:1998gyp}.

In this work, we will make use of the numerical conformal bootstrap using $\mathrm{SL}(2,\mathbb{R})$ (global) conformal blocks, i.e.\ the same bootstrap formalism which has led to the precision results in $d=3$. See~\citep{Chester:2019wfx} for lecture notes. That is, we will be working in terms of \emph{quasi-primary} operators and only considering the $\mathrm{sl}(2,\mathbb{R})$ sub-algebra of the Virasoro algebra spanned by $L_1$, $L_0$ and $L_{-1}$. 
In practice, we impose self-consistency, i.e.\ crossing symmetry, in various 4-point correlators involving three of the lowest spin-0 operators in the spectrum, in terms of scaling dimension. These in our notation will be called $\phi$, $X$ and $Z_{3}$ and have dimensions respectively $\Delta_\phi = 2/15$, $\Delta_X =4/5$, and $\Delta_{Z_3} =6/15$ for $n=3$ and $Q=3$ at the decoupled fixed point, i.e.\ for $g=0$ in Eq.~\eqref{lagrangian}. The second lowest operator in the theory is also constrained by our analysis, where it enters as an exchanged operator. We call this operator $Z$ and it has dimension $\Delta_{Z}=4/15$ at the decoupled fixed point. 
For all four operators, we compare to available perturbative and non-perturbative results from the literature, which include data from Monte Carlo simulations~\citep{Dotsenko:1998gyp} as well as transfer matrices~\citep{Dotsenko:1998gyp,Dotsenko:2001cct}. In particular, we obtain an allowed peninsula in the parameter space spanned by $(\Delta_\phi, \Delta_Z, \Delta_{Z_3})$ which nicely includes the earlier predictions of~\citep{Dotsenko:1998gyp}. This indicates that the conformal bootstrap is indeed strong enough to constrain the low-lying spectrum of these theories.
Notice that because the singlet in the theory has a dimension $\Delta_S \simeq 1.2$,\footnote{This value can also be understood qualitatively from the large-$n$ expansion~\citep{Binder:2021vep}. In the large-$n$ limit the lowest lying scalar, i.e.\ the auxiliary field, has the shadow dimension of the lowest lying singlet in the decoupled theory, i.e.\ $\Delta_S \rightarrow d - \Delta_S\ap{dec} =2-4/5=1.2$. See later Table~\ref{tab:epsexp} for perturbative and non-perturbative estimates from~\citep{Dotsenko:1998gyp}. That being said, we will see operators, such as $Z_3$, which acquire particularly large anomalous dimensions. Hence large-$n$ intuition should be used in moderation.} this differs from the usual situation regarding most theories studied in $d=3$ where the first scalar singlet is typically one of the lowest-lying operators in the theory. 
In practice, this means that in order to access the low-lying part of the spectrum we will need to study a much more numerically costly set of sum rules. This is because the correlators of the non-singlet operators have more tensor structures and thus the crossing symmetry constraint produces more sum rules. \\

With respect to conventions throughout this work, a four-point correlator denoted as $\langle O_1 O_2 O_3 O_3\rangle$ will always imply that the OPE is to be taken between the first two and last two operators respectively. Also, from this point forth, the phrase ``global symmetry'', will refer to the global ``flavour'' symmetry i.e.\ $S_n \ltimes (S_Q)^n$, and {\bf not} the global part of the Virasoro algebra. The latter of which is included in what we will call the ``spacetime symmetry''. Lastly, when discussing the lowest-lying operators in the theory, we do not refer to the identity operator, so that e.g.\ by ``lowest-lying singlet'' we refer to a non-trivial field. \\

The paper is organised as follows. In Section \ref{sec:introGroupTheory} we outline all the necessary group theory for building operators in a given representation and deriving the sum rules. This is done in a way appropriate for generic $n$ and $Q$. In Section \ref{sec:PreviousResults} we review previous results from the literature which we will make contact to in the rest of the work. Then, in Section \ref{sec:OurResults} we present the numerical constraints we have derived, and conclude with Section \ref{sec:Discussion}, where we discuss various directions we expect to prove fruitful in future studies. In order to aid readability, we have relegated further technical details to Appendices \ref{sec:SumRules}, \ref{numericalparameters}, \ref{AppQuasiPrimaries}, \ref{sec:CharTable} and \ref{discreteformulas}. These can be skipped on a first read-through. 

\section{Elements of Group and Field Theory}
\label{sec:introGroupTheory}
In this work, we will follow the naming conventions of~\citep{Bednyakov:2023lfj}, which dealt with the symmetry group $S_n \ltimes (\mathbb{Z}_2)^n$ describing theories with hypercubic symmetry. The group we will cover here is a special case of a wreath product $ G \wr S_{n} = S_n \ltimes G^n$, where $G=S_3$ or more generally $G=S_Q$. This class of theories, their group theory, as well as their tensor structures as appropriate to the bootstrap have been studied in~\citep{Kousvos:2021rar}, from where we will also borrow. Note that while $S_3 \ltimes (S_3)^3$ is itself a finite group and its character table can be obtained with {\bf GAP}~\citep{GAP4}\footnote{\href{https://www.gap-system.org/}{https://www.gap-system.org/}} (see Table~\ref{tab:character-table}), the way we will treat the group theory in this work will allow us to treat the groups $S_n \ltimes (S_Q)^n$ for arbitrary values of $Q$ and $n$, including non-integer, as well as large parameter limits such as large-$n$ which would be infeasible with the software currently available.\footnote{Let us also remark that while {\bf autoboot}~\citep{Go:2019lke} can in principle treat all finite groups, in practice the order of the group $S_3 \ltimes (S_3)^3$ is so large that its usage is impractical if not infeasible with the current implementation.} It will also enable us to study complicated mixed correlators in theories, such as the $S_n \ltimes O(m)^n$ symmetric multi-scalar theories of the $d=4-\varepsilon$ dimension expansion, in future work with only small modifications. See~\citep{Pelissetto:2000ek,Stergiou:2019dcv,Henriksson:2021lwn,Kousvos:2021rar} for more information on these theories.

\subsection{The Defining Representation}
One intuitive way to start discussing the group theory, which will also make direct contact with the field content of the CFT, is to consider a field transforming in the ``defining'', representation of $S_n \ltimes (S_Q)^n$ dubbed
\begin{equation}
\phi^a_i \, .
\label{fielddefinition}
\end{equation}
Here the upper index is acted on by $S_n$, running from $a=1$ to $n$, and it labels the copy/replica to which the field belongs. Whereas the lower index $i$ runs from $1$ to $Q$, and transforms in the defining representation of $S_Q$. This corresponds to the field with which the Lagrangian is built in the perturbative limit. The dimension of the irrep carried by $\phi^a_i$ under $S_n \ltimes (S_Q)^n$ is $\dim(\phi) = n (Q-1)$, which is $n$ times the dimension of the defining representation of $S_Q$, since we have $n$ replicas of it. We will label this representation as $V$ and $\phi$ interchangeably.

For a generic wreath product $G \wr S_{n}$, we can build ``simple'' representations in a similar way. We consider $n$ copies of a field, denoted collectively as $O_{i}^{a}$, which transforms in each copy under an irrep $\rho$ of $G$, while the copy index can be permuted under $S_{n}$:
\begin{equation}
    O_{i}^{a} \xrightarrow{(\vec{g}; \pi)} \sum_{b=1}^{n}\sum_{i=1}^{\dim(\rho)} \rho(g_a)_{ij} \, P^{ab} \, O_{j}^{b} \, .
\end{equation}
Here $(\vec{g}; \pi) = ((g_{1}, \ldots, g_{n}); \pi) \in G \wr S_{n}$ (namely $g_{a} \in G$ and $ \pi \in S_{n}$) and $P$ is the permutation matrix associated to $\pi$. This representation is irreducible and has dimension ${\dim(O)=n\,\dim(\rho)}$ unless $\rho$ is the trivial representation of $G$. In the latter case the field $O^{a}$ is effectively a vector of dimension $n$ which transforms under $S_{n}$ only. This representation is reducible and can be split into a singlet and a vector representation of $S_{n}$:
\begin{align}
    S &= \sum_{a=1}^{n} O^{a} \, , \\
    X^{a} &= O^{a} - \frac{1}{n}\sum_{b=1}^{n} O^{b} \, .
\end{align}
of dimension $\dim(S) = 1$ and $\dim(X) = n-1$ respectively. For a mathematically precise treatment of representations of wreath products see~\citep{James1981}.

\subsection{Decomposing the Product of Two Defining Representations}

The rest of the representations in the group can now be built by taking products of $\phi$'s and decomposing them into irreducible components. See Table~\ref{tab:irreps} for the complete list of irreps when $n=3, Q=3$. Let us work out the decomposition of the product between two $\phi$'s to give an explicit example in the $Q=3$ case. This has also been outlined for $S_n \ltimes O(m)^n$ in~\citep{Kousvos:2021rar}, of which the groups we are considering are subgroups. We have 

\begin{equation}
\phi^a_i \otimes \phi^{\prime b}_j \sim \delta^{ab} \delta_{ij} S + \delta_{ij} X^{ab} + A^{ab}_{ij} + C^{abc}_{ijk} \phi^c_k +Z^{ab}_{ij} + B^{ab}_{ij} \, ,
\label{productoftwophis}
\end{equation}
where the decomposition rules are as follows: 
\begin{itemize}
    \item if $a=b$ we decompose the lower indices onto irreps of $S_3$ (or more generally $S_Q$);
    \item if $a \neq b$ then we symmetrize and antisymmetrize the $\phi$ fields, one of which carries a prime so that the antisymmetric combination doesn't vanish identically.
\end{itemize}
The tensors $C^{abc}_{ijk}$ are defined as
\begin{equation}
C^{abc}_{ijk} = \delta^{abc} c_{ijk}\,,
\end{equation}
where $c_{ijk}$ is proportional to the tensor structure of a three-point function of fields transforming in the defining representation of $S_3$, $\langle O_i O_j O_k \rangle$.

The first four irreps on the right hand side of Eq.~\eqref{productoftwophis}, ($S$, $X^{ab}$, $A^{ab}_{ij}$, $\phi^c_k$), only appear when $a=b$, whereas $Z^{ab}_{ij}$ and $B^{ab}_{ij}$ only appear when $a\neq b$. These can be expressed as 
\begin{gather}
    S= \delta^{ab}\delta_{ij} \phi^a_i \phi^{\prime b}_j \, ,\label{Soperator} \\
    X^{ab} = \left(\delta^{abcd}-\frac{1}{n}\delta^{ab}\delta^{cd}\right)\delta_{ij} \phi^c_i \phi^{\prime d}_j\,, \label{Xoperator} \\
    A^{ab}_{ij} = \delta^{abcd} P^A_{ijkl} \phi^c_i \phi^{\prime d}_j \,,\label{Aoperator} \\
    Z^{ab}_{ij} =\left( (\delta^{ac}\delta^{bd}-\delta^{abcd})\delta_{ik}\delta_{jl}+(\delta^{ad}\delta^{bc}- \delta^{abcd})\delta_{il}\delta_{jk} \right) \phi^c_i \phi^{\prime d}_j \,,\label{Zoperator} \\
    B^{ab}_{ij} =\left( (\delta^{ac}\delta^{bd}-\delta^{abcd})\delta_{ik}\delta_{jl}-(\delta^{ad}\delta^{bc}- \delta^{abcd})\delta_{il}\delta_{jk} \right) \phi^c_i \phi^{\prime d}_j \,,\label{Boperator}
\end{gather}
where $P^A$ is the tensor that projects a product of two fields onto the rank-two antisymmetric representation of $S_3$ (or $S_Q$), see Appendix~\ref{singcorrphi}. The projectors typically contain additional factors of $\omega_i$, which satisfy $\sum_i \omega_i = Q$, and each component of the vector $\omega_i$ is equal to one. Hence in the above relations, they can be omitted, since $\sum_i \omega_i \phi^a_i =\sum_i \phi^a_i =0$, where the last equality follows from the definition of the $Q-1$ dimensional defining irrep of $S_Q$.

\begin{table}[t]
    \centering
    \caption{Some low-lying irreps of $S_n \ltimes (S_3)^n$. The list is complete for $n=3$. The dimension of each representation is reported for $n=3$, but each irrep in this table also exists for $n>3$.}\label{tab:irreps}
    \begin{tabular}{cc}
      \toprule
      Symbol & dimension\\ \midrule
      $S$   & $1$\\
      $\overline{AAA}$    & $1$\\
      $\overline{XX}$    & $1$\\
      $AAA$    & $1$\\
      $\overline{AA}A$    & $2$\\
      $X$    & $2$\\
      $XA$    & $3$\\
      $\overline{AA}$   & $3$\\
      $A$    &$3$\\
      $AA$    & $3$\\
      $V\overline{AA}$ & $6$\\
      \bottomrule
    \end{tabular}
    \hspace{1em}
    \begin{tabular}{cc}
      \toprule
      Symbol & dimension\\ \midrule
      $XV$ & $6$\\
      $VAA$ &$6$\\
      $\phi = V$ &$6$\\
      $ B_3 =\overline{VVV}$ &$8$\\
      $Z_3=VVV $ &$8$\\
      $VA$ &$12$\\
      $ AB=A\overline{VV}$ &$12$\\
      $ B=\overline{VV}$ &$12$\\
      $ AZ=AVV $ &$12$\\
      $ Z=VV $ &$12$\\
      $ VB=V\overline{VV}$ & $16$\\
      \bottomrule
    \end{tabular}
\end{table}

For a generic wreath product $G \wr S_{n}$, we can decompose the tensor product of two ``simple'' irreps $O^{a}_{i} \otimes \tilde{O}^{b}_{j}$ (transforming in the $\rho$ and $\tilde{\rho}$ irreps under $G$) with a similar logic.
\begin{itemize}
    \item If $a = b$ we decompose the lower indices into irreps of $G$: $\rho \otimes \tilde{\rho} = r_{1} \oplus \cdots \oplus r_{m}$. Each of them is irreducible, unless $r_{k}$ is the trivial representation;  in the latter case, as previously discussed, the representation is reducible and can be decomposed in $S \oplus X$. This is what happens in Eq.~\eqref{productoftwophis}: the tensor product of two defining representations of $S_{3}$ decomposes into singlet ($S^{a}$), defining ($\phi^{a}_{i}$) and antisymmetric ($A^{a}_{ij}$) representations; the ``replicated'' singlet is then further decomposed into $S$ and $X$ components.
    \item If $a \neq b$ and $\rho \neq \tilde{\rho}$ the representation is irreducible.
    \item If $a \neq b$ and $\rho = \tilde{\rho}$ the representation is reducible under the action of $S_{n}$ and can be decomposed into a symmetric and antisymmetric part. In Eq.~\eqref{productoftwophis} these are the $Z^{ab}_{ij}$ and $B^{ab}_{ij}$ representations.
\end{itemize}

\subsection{Higher Representations}
\label{compositeirreps}
While carrying out the above procedure of decomposing an arbitrary product of $\phi$'s would prove to be highly impractical very quickly, it is easy to infer (and check with the help of the character table), the entire representation content of the theory. We restrict to $S_3 \ltimes (S_3)^3$, and $S_3 \ltimes (S_Q)^3$, for practical reasons.

As an example, let us work out all the representations that can be built using the three replicas of the antisymmetric representation, ($A^1$, $A^2$, $A^3$), of each of the three $S_Q$ factors included in $S_3 \ltimes (S_Q)^3$. We can take the totally symmetric product of three of these $A^{(1}A^2 A^{3)}$, or of two $A^{(1}A^{2)}$ or of just one of them $A^1$. These are precisely the irreps $AAA$, $AA$ and $A$ in Table~\ref{tab:irreps}. Similarly, one can also antisymmetrize the replicas of $A$. One has $A^{[1}A^2 A^{3]}$ and $A^{[1}A^{2]}$, which are denoted as $\overline{AAA}$, and $\overline{AA}$ in Table~\ref{tab:irreps}. Notice that whenever factors of a representation are antisymmetrized we place a bar over them, this is the notation used in~\citep{Bednyakov:2023lfj}. Beyond totally symmetric and totally antisymmetric combinations of $A$'s, we can form combinations that form other Young Tableaux of $S_3$, which produces $\overline{AA}A$. Lastly, in cases where all three replicas have not been exhausted, which in this example are $AA$, $A$ and $\overline{AA}$, we can also adjoin to them another representation to create a new irrep. For example $AA$ can be adjoined with $V$ to produce $V^1 A^{(2}A^{3)}$, which is $VAA$ in Table~\ref{tab:irreps}.

Together with the representation $X$ from earlier in the product of two fields, these considerations explain all representations in Table~\ref{tab:irreps}. In the following $Z$, $B$, $Z_3$, $B_3$ and $VB$ are simply shorthand for $VV$, $\overline{VV}$, $VVV$, $\overline{VVV}$, $V\overline{VV}$. This understanding of the group theory in terms of  ``composite'' representations, will provide important intuition later when we consider gap assumptions to be imposed on admissible spectra. In the case of $S_n \ltimes (S_3)^n$, for example, the scaling dimensions of all operators are given in terms of the ones of the single critical $3$-state Potts model (see Appendix~\ref{AppQuasiPrimaries}) in the decoupled or large-$n$ limits.

\section{Previous Results}
\label{sec:PreviousResults}

We point the reader to~\citep{Dotsenko:1998gyp} and references therein for a number of useful resources. To the best of our knowledge, the theories under consideration in this paper have been studied using perturbation theory, transfer matrices and Monte Carlo simulations.

With respect to perturbation theory, while one may in principle perform an expansion continuing in flavours, an expansion at large number of copies~\citep{Binder:2021vep}, or a fixed dimension expansion~\citep{Serone:2018gjo}, only the first has been carried out explicitly~\citep{Ludwig:1987rk,Ludwig:1989rj,Dotsenko:1994im,Dotsenko:1994sy}.\footnote{Additional references include \citep{Cardy_1996,Pujol_1996,Simon_1998,Lewis:1998he,Lewis_1998,Dotsenko:1997wf}.} In this case, one considers an expansion of $S_n \ltimes (S_{2+\varepsilon})^n$ around $\varepsilon$ infinitesimal, where the flow triggered by the ``energy-energy'' interaction of Eq.~\eqref{lagrangian} becomes controlled. In other words, one perturbs around $n$ replicas of the Ising model ($Q = 2$). A caveat in this case is that for $Q=2$ a lot of representations do not appear: all irreps in Table~\ref{tab:irreps} that include a factor of $A$ cease to exist, and thus data for them cannot be extracted.

The perturbative results are complemented by numerous transfer matrix computations. In this approach, one computes an approximation to the partition function of the lattice model on the cylinder, from which it is possible to extract the eigen-energies. These in turn are proportional to the scaling dimensions of the theory in the continuum limit. To this end, in~\citep{Dotsenko:1998gyp} the authors calculated the central charge and the scaling dimensions of what we call $\phi$, $Z$, $Z_3$ and $S$. More concretely, they calculated these for the leading spin-$0$ operator in each representation. Furthermore, in the singlet sector $S$ they computed the first 8 eigen-energies, which correspond to both $SO(2)$ scalars and spinning operators (e.g.\ the stress tensor), see Table 17 in that work. Consequently, the scaling dimension of the leading spin-$0$ $X$ operator was computed in~\citep{Dotsenko:2001cct}.

Lastly, in~\citep{Dotsenko:1998gyp}, the authors also performed a Monte Carlo study in order to corroborate the existence of a critical fixed point. They computed the decay of two-point functions with distance and found a behaviour in agreement with criticality. This led to the extraction of the scaling dimension for operators in the $\phi$, $X$ and $S$ representations. These were also found to be in good agreement with the transfer matrix results.

\begin{table}[t]
\caption{Dimensions of operators from~\citep{Dotsenko:1998gyp} (and~\citep{Dotsenko:2001cct} in the case of Transfer Matrices for $X$). For the original computations in each case see \citep{Dotsenko:1998gyp} and references therein. The results are reported in conventions where $\varepsilon=2/15$ corresponds to replicas of $3$-state Potts models and $\varepsilon=1/3$ to $4$-state Potts models, see below equation 3.2 in~\citep{Dotsenko:1998gyp}. Some transfer matrix results are not reported with error bars in the original work, hence we also omit them. Thus the choice of digits does not imply any error bar unless explicitly stated. Primes denote subleading operators and $\alpha$ and $\mathcal{F}$ are defined as $\alpha = 33 - 29 \sqrt{3} \pi/3$ and $\mathcal{F}=2\Gamma\left(-\frac{2}{3}\right)^2 \, \Gamma\left(\frac{1}{6}\right)^2 / \left(\Gamma\left(-\frac{1}{3}\right)^2 \, \Gamma\left(-\frac{1}{6}\right)^2\right)$ respectively. The perturbative results for the $S$ and $X$ representations are reported at $n=3$. The factors $\Delta_\sigma (\varepsilon)$ and $\Delta_\epsilon (\varepsilon)$ are the decoupled theory dimensions and are equal to $2/15$ and $4/5$ respectively at $Q=3$.}\label{tab:epsexp}
\aboverulesep=0ex
\belowrulesep=0ex
\resizebox{\textwidth}{!}{
\begin{NiceTabular}{c|cccc}[cell-space-limits=2pt]
\toprule 
Operator & Expression & $\varepsilon = 2/15$ & Tr.Matrix & M.Carlo
\\
\midrule
$S$   & $\Delta_\epsilon (\varepsilon) + 6 \varepsilon - 9 \varepsilon^2$ & $1.44$ & $1.27190$& $1.27 \pm 0.13$
\\
$S^\prime$     & $2\Delta_\epsilon (\varepsilon) + 3 \varepsilon + \frac{9}{2} \varepsilon^2$&  $2.08$   & $\sim 2.1$& 
\\
$S^{\prime \prime}$    & $3\Delta_\epsilon (\varepsilon) - \frac{27}{4} \varepsilon^2$&  $2.28$  & $\sim 2.4$& 
\\
$X$    & $\Delta_\epsilon (\varepsilon) - 3\varepsilon + \frac{9}{2} \varepsilon^2$&  $0.48$  & $0.63 \pm 0.03$ & $0.63 \pm 0.04$
\\
$X^\prime$    & $2\Delta_\epsilon (\varepsilon) - \frac{3}{2}\varepsilon - 9 \varepsilon^2$& $1.24$ & & 
\\
$Z$    & $2\Delta_\sigma (\varepsilon) + \frac{3 \varepsilon}{4(n-2)} \left(1 - 
     \frac{3\varepsilon}{n-2}\left((n - 2) \log{2} + \frac{11}{12}\right)\right)$& $0.30227$  & $0.30960$& 
\\
$Z_3$    & $3\Delta_\sigma (\varepsilon) + \frac{9\varepsilon}{4(n-2)} \left(1 - 
     \frac{3\varepsilon}{n-2}\left((n - 2) \log{2} + \frac{11}{12} + \frac{\alpha}{24}\right) \right)$ &  $0.60482$  & $0.59222$ & 
\\
$\phi = V$    &$\Delta_\sigma (\varepsilon)-\frac{27}{32}\frac{n-1}{(n-2)^2}\mathcal{F}\varepsilon^3$ &$0.12805$    &$0.13162$ & $0.13 \pm 0.03$
\\
\bottomrule
\end{NiceTabular}}
\end{table}

\section{Results}
\label{sec:OurResults}
We will study a number of different four-point correlator systems on which we will impose self-consistency (i.e.\ crossing symmetry). We remind the reader of the convention where $\langle O_1 O_2 O_3 O_4 \rangle$ implies that the given four-point correlator is computed by taking the OPE of $O_1 \times O_2$ and $O_3 \times O_4$. Hence, an equation of the form
\[
    \langle O_1 O_2 O_3 O_4 \rangle = \langle O_3 O_2 O_1 O_4 \rangle
\]
will simply imply that the OPEs are taken in a different order, and is shorthand for
\[
    \langle O_1(x_1) O_2(x_2) O_3(x_3) O_4(x_4) \rangle = \langle O_3(x_3) O_2(x_2) O_1(x_1) O_4(x_4) \rangle \, .
\]
With this in mind, let us specify the correlator systems we will study. 

These are:
\begin{itemize}
    \item the system involving only $\phi^a_i$, which will be referred to as the single correlator $\phi$ bootstrap:
\begin{equation}
\langle \phi^a_i \phi^b_j \phi^c_k \phi^d_l \rangle = \langle \phi^c_k \phi^b_j \phi^a_i \phi^d_l \rangle\,;
\label{singlecrossing}
\end{equation}

\item the mixed system involving both $\phi^a_i$ and $X^{ab}$, which will be referred to as the mixed correlator $\phi - X$ bootstrap:
\begin{equation}
\begin{aligned}
\langle \phi^a_i \phi^b_j \phi^c_k \phi^d_l \rangle &= \langle \phi^c_k \phi^b_j \phi^a_i \phi^d_l \rangle \,,\\
\langle X^{ab} X^{cd} X^{ef} X^{gh}\rangle &= \langle X^{ef} X^{cd} X^{ab} X^{gh}\rangle \,, \\
\langle \phi^a_i  X^{cd} \phi^b_j X^{ef} \rangle &= \langle \phi^b_j  X^{cd} \phi^a_i X^{ef} \rangle \,,\\
\langle \phi^a_i  \phi^b_j X^{cd}  X^{ef} \rangle &= \langle  X^{cd} \phi^b_j  \phi^a_i X^{ef} \rangle\,;
\end{aligned}
\label{crossingPhiX}
\end{equation}

\item the the system involving only ${Z_3}^{\lsp abc}_{\lsp ijk}$, which will be referred to as the single correlator $Z_3$ bootstrap:
\begin{equation}
\langle {Z_3}^{\lsp a_1b_1c_1}_{\lsp ijk}{Z_3}^{\lsp a_2b_2c_2}_{\lsp lmn}{Z_3}^{\lsp a_3b_3c_3}_{\lsp opr}{Z_3}^{\lsp a_4b_4c_4}_{\lsp stu} \rangle = \langle {Z_3}^{\lsp a_3b_3c_3}_{\lsp opr} {Z_3}^{\lsp a_2b_2c_2}_{\lsp lmn}{Z_3}^{\lsp a_1b_1c_1}_{\lsp ijk}{Z_3}^{\lsp a_4b_4c_4}_{\lsp stu} \rangle \,;
\label{crossingZ3single}
\end{equation}

\item  the mixed system involving both $\phi^a_i$ and ${Z_3}^{\lsp abc}_{\lsp ijk}$, which will be referred to as the mixed correlator $\phi - Z_{3}$ bootstrap:
\begin{equation}
\begin{aligned}
\langle \phi^a_i \phi^b_j \phi^c_k \phi^d_l \rangle &= \langle \phi^c_k \phi^b_j \phi^a_i \phi^d_l \rangle \,, \\
\langle \phi^a_i  {Z_3}^{\lsp bcd}_{\lsp jkl} \phi^e_{m} {Z_3}^{\lsp fgh}_{\lsp nop}\rangle &= \langle \phi^e_m  {Z_3}^{\lsp bcd}_{\lsp jkl} \phi^a_{i} {Z_3}^{\lsp fgh}_{\lsp nop}\rangle \,, \\
\langle \phi^a_i \phi^e_m {Z_3}^{\lsp bcd}_{\lsp jkl} {Z_3}^{\lsp fgh}_{nop}\rangle &= \langle {Z_3}^{\lsp bcd}_{\lsp jkl}  \phi^e_m  \phi^a_i {Z_3}^{\lsp fgh}_{nop} \rangle \,,\\
\langle {Z_3}^{\lsp a_1b_1c_1}_{\lsp ijk}{Z_3}^{\lsp a_2b_2c_2}_{\lsp lmn}{Z_3}^{\lsp a_3b_3c_3}_{\lsp opr}{Z_3}^{\lsp a_4b_4c_4}_{\lsp stu} \rangle &= \langle {Z_3}^{\lsp a_3b_3c_3}_{\lsp opr} {Z_3}^{\lsp a_2b_2c_2}_{\lsp lmn}{Z_3}^{\lsp a_1b_1c_1}_{\lsp ijk}{Z_3}^{\lsp a_4b_4c_4}_{\lsp stu} \rangle \,.
\end{aligned}
\label{crossingPhiZ3}
\end{equation}

\end{itemize}

As a heavy-handed indicator of numerical cost, we mention that for $n=3,Q=3$ the above systems give $6$, $15$, $10$ and $28$ sum rules respectively. 

Each of these systems offers various advantages and disadvantages. The $\phi$ single correlator bootstrap has a small number of sum rules, hence it is numerically less costly and also has the lightest operator in the theory as an external. It is however less constraining than the mixed correlator systems we will study. The $\phi - X$ system contains fewer sum rules than the $\phi - Z_{3}$ system but does not allow us (as we will see) to easily segregate the coupled and the decoupled fixed points in parameter space. On the other hand, while a potential $\phi - Z$ mixed correlator system would involve the two lightest operators in the theory, and would presumably be the most constraining, it is considerably more costly numerically than the $\phi - Z_{3}$ system. In addition to this, the first $Z$ operator at spin-$0$ has a similar dimension at both the coupled and decoupled fixed points, making it hard to distinguish the two fixed points in the numerics.\footnote{Whereas the first spin-$0$ $Z_3$ operator obtains a rather large anomalous dimension at the coupled fixed point. See Table~\ref{tab:epsexp}.} Hence, we omit an analysis of the $\phi - Z$ system in the present work.

Note that in addition to the correlator systems discussed above, we also performed some preliminary tests with the mixed correlator $\phi - \phi^\prime$ bootstrap
\begin{equation}
\begin{aligned}
\langle \phi^a_i \phi^b_j \phi^c_k \phi^d_l \rangle &= \langle \phi^c_k \phi^b_j \phi^a_i \phi^d_l \rangle \,,\\
\langle \phi^{\prime a}_i \phi^{\prime b}_j \phi^{\prime c}_k \phi^{\prime d}_l \rangle &= \langle \phi^{\prime c}_k \phi^{\prime b}_j \phi^{\prime a}_i \phi^{\prime d}_l \rangle  \,,\\
\langle \phi^{ a}_i \phi^{\prime b}_j \phi^{ c}_k \phi^{\prime d}_l \rangle &= \langle \phi^{ c}_k \phi^{\prime b}_j \phi^{ a}_i \phi^{\prime d}_l \rangle \,, \\
\langle \phi^{ a}_i \phi^{ b}_j \phi^{\prime c}_k \phi^{\prime d}_l \rangle &= \langle \phi^{\prime c}_k \phi^{ b}_j \phi^{ a}_i \phi^{\prime d}_l  \rangle \, .
\end{aligned}
\label{crossingPhiPhiprime}
\end{equation}
The rationale behind this system is that it exchanges the $S$ and $V$ representations at spin-$1$, and the $A$ representation at spin-$0$. At least at the decoupled fixed point, whose scaling dimensions are determined through the $3$-state Potts model, these operators possess large gaps (see Appendix~\ref{AppQuasiPrimaries}). However, we found this system less constraining than the ones we present in this work. While there may be many reasons for this, one obvious guess is that since $\Delta_{\phi^\prime}=4/3 $ at the decoupled fixed point (and presumably somewhat close in the coupled one), it is significantly heavier than all the other operators we study as externals. 
In particular, we observed the ``no mixing'' problem,\footnote{Described in the following lectures \href{https://perimeterinstitute.ca/events/mini-course-numerical-conformal-bootstrap}{https://perimeterinstitute.ca/events/mini-course-numerical-conformal-bootstrap}. Essentially, the bootstrap functional reduces to that of the single correlator system by putting all components involving the mixed parts to what is practically zero. See the slides at \href{http://scgp.stonybrook.edu/video_portal/video.php?id=4327}{http://scgp.stonybrook.edu/video\_portal/video.php?id=4327}, slides $11$ and $12$ in particular.} which is not observed in the other systems we study, at least in the vicinity of the fixed point.

Let us finally mention that we also experimented with a mixed correlator $\phi - S$ bootstrap
\begin{equation}
\begin{aligned}
\langle \phi^a_i \phi^b_j \phi^c_k \phi^d_l \rangle &= \langle \phi^c_k \phi^b_j \phi^a_i \phi^d_l \rangle \,, \\
\langle S S S S\rangle &= \langle S S S S\rangle  \,,\\
\langle \phi^a_i  S \phi^b_j S \rangle &= \langle \phi^b_j  S \phi^a_i S \rangle \,,\\
\langle \phi^a_i  \phi^b_j S S \rangle &= \langle  S \phi^b_j  \phi^a_i S \rangle \,,
\end{aligned}
\label{crossingPhiS}
\end{equation}
since this is the ``cheapest'' mixed correlator system one might consider. We found this system to also be less constraining than the ones we present in this work. As for the $\phi-\phi'$ system, the most probable reason is the fact that the singlet is expected to have $\Delta_{S} \simeq 1.2$ and is thus a relatively heavy operator. \\

The sum rules we derive have the schematic form
\begin{equation}
    \mathcal{V}^{\Delta\ped{ext}}_{\id} + \vec{\theta} \cdot \mathcal{V}^{\Delta\ped{ext}}_{\theta} \cdot \vec{\theta} + \sum_{O= (R,\Delta,\ell)} \vec{\lambda}_{O} \cdot \mathcal{V}^{\Delta\ped{ext}}_{O} \cdot \vec{\lambda}_{O} = 0
\end{equation}
where $\mathcal{V}$ are vectors made out of convolved conformal blocks (see Appendix~\ref{apx:sumrules} for more details). $\mathcal{V}^{\Delta\ped{ext}}_{O}$ are generically vectors of matrices in mixed correlator systems. In the sum, $R$ labels the irrep under $S_{n} \ltimes (S_{Q})^{n}$. The vector $\vec{\lambda}_{O}$ is a collection OPE coefficients between a pair of externals and the exchanged operator $O$. The identity is always exchanged in our sum rules and we separate it explicitly ${\mathcal{V}^{\Delta\ped{ext}}_{\id} = \vec{\lambda}_{\id} \cdot \mathcal{V}^{\Delta\ped{ext}}_{\id} \cdot \vec{\lambda}_{\id}}$. Moreover, since in the present work the external operators also appear as exchanged (see Table~\ref{tab:opes}), we explicitly separate them in the term $\mathcal{V}^{\Delta\ped{ext}}_{\theta} = \mathcal{V}^{\Delta\ped{ext}}_{O = \mathrm{ext}}$ and $\vec{\theta} = \vec{\lambda}_{O = \mathrm{ext}}$. 

The numerical bootstrap approach rules out putative CFT spectra by looking for a linear functional $\alpha$ satisfying $\alpha(\mathcal{V}_{\id}) = 1$ and certain positivity conditions on $\alpha(\mathcal{V}_{O})$ which depend on the gap assumptions on the spectrum. If the functional exists the assumed gaps are inconsistent with crossing symmetry. When looking for the optimal bound $\Delta^{*}$ on the dimension of the lightest operator in the $R^{*}$ representation and of spin $\ell^{*}$, we look for functionals satisfying schematically the following conditions
\begin{equation}
    \begin{aligned}
        \alpha(\mathcal{V}_{\id}^{\Delta\ped{ext}}) &= 1 \,,\\
        \alpha(\mathcal{V}_{\theta}^{\Delta\ped{ext}}) &\geq 0 \,,\\
        \alpha(\mathcal{V}_{R\ped{ext},\Delta,\ell}^{\Delta\ped{ext}}) &\geq 0 \qquad && \forall \ell; \, \Delta \geq \Delta\ped{gap} \,,\\
        \alpha(\mathcal{V}_{R^{*},\Delta,\ell^{*}}^{\Delta\ped{ext}}) &\geq 0 \qquad && \forall \Delta \geq \Delta^{*} \,,\\
        \alpha(\mathcal{V}_{R,\Delta,\ell}^{\Delta\ped{ext}}) &\geq 0 \qquad && \forall \ell \neq \ell^{*}; \, \forall R \neq R^{*},R\ped{ext}; \,  \forall \Delta \geq \Delta\ped{gap}  \,.\\
    \end{aligned}
\end{equation}
Here $\Delta\ped{gap} = \Delta\ped{gap}(R,\ell)$ is a set of fixed gap assumptions (see Table~\ref{tab:gaps} and subsequent comments). As already mentioned the externals are always exchanged so we explicitly look for a functional $\alpha(\mathcal{V}_{\theta}^{\Delta\ped{ext}}) \geq 0$ and impose a fixed gap for the next operator in that sector. The problem is parametric in $\Delta\ped{ext}$ and $\Delta^{*}$, which are thus scanned over to produce the plots reported in this work.

In some cases, we look for the allowed region for the dimension of the lowest-lying operator $O_{R^{*},\Delta^{*},\ell^{*}}$ by putting it as isolated in the spectrum and imposing a fixed gap above it (in a similar way to what we do for external operators). In this case the positivity condition reads
\begin{equation}
    \begin{aligned}
        \alpha(\mathcal{V}_{\id}^{\Delta\ped{ext}}) &= 1 \,,\\
        \alpha(\mathcal{V}_{\theta}^{\Delta\ped{ext}}) &\geq 0 \,,\\
        \alpha(\mathcal{V}_{R\ped{ext},\Delta,\ell}^{\Delta\ped{ext}}) &\geq 0 \qquad &&  \forall \ell; \, \forall \Delta \geq \Delta\ped{gap} \,,\\
        \alpha(\mathcal{V}_{R^{*},\Delta^{*},\ell^{*}}^{\Delta\ped{ext}}) &\geq 0 \,,\\
        \alpha(\mathcal{V}_{R^{*},\Delta,\ell^{*}}^{\Delta\ped{ext}}) &\geq 0 && \forall \Delta \geq \Delta\ped{gap}\,,\\
        \alpha(\mathcal{V}_{R,\Delta,\ell}^{\Delta\ped{ext}}) &\geq 0 \qquad &&  \forall \ell \neq \ell^{*}; \, \forall R \neq R^{*},R\ped{ext}; \, \forall \Delta \geq \Delta\ped{gap}\,.\\
    \end{aligned}
\end{equation}
The search for the functional is done again parametrically in $(\Delta\ped{ext},\Delta^{*})$. \\

In favour of readability we have relegated the resulting sum rules from the above crossing equations, \eqref{singlecrossing} to \eqref{crossingPhiZ3}, and their derivation to a number of appendices, namely~\ref{singcorrphi} through~\ref{phiz3sumrules}. Tables~\ref{tab:opes} and~\ref{tab:gaps} collect all OPEs and gap assumptions relevant to the figures that will follow.  If a gap assumption is mentioned in the caption of a figure, which concerns an irrep already covered in Table~\ref{tab:gaps}, {\bf then the assumption in the caption replaces the one in Table~\ref{tab:gaps}}. Also, when the dimension of an operator is being bounded, we don't impose the corresponding gap for that operator found in the table. This choice was made to shorten the presentation and make the present manuscript more readable. 

All figures that will follow concern $n=3,Q=3$, which the exception of a sub-figure in Fig~\ref{fig:Zspin0singlecorrlargeN} computed at $n=100$ and $Q=3$. In all plots, shaded areas present the admissible parameter space, surrounded by the excluded region. The blue dot denotes the coupled replica fixed point (transfer matrix values in Table~\ref{tab:epsexp}), whereas the magenta dot the fixed point of decoupled replicas.
\begin{table}[t]
\caption{OPEs relevant to the present work for $n=3,Q=3$. While we use $\phi \times \phi$ at $n=100$ later on, its form remains as reported in this table (same exchanged irreps). A plus (resp.\ minus) subscript on a representation signifies it is exchanged with even (resp.\ odd) spins. OPEs with non-identical external irreps exchange both spins.}\label{tab:opes}
\centering
\renewcommand{\arraystretch}{0.8}
\resizebox{\textwidth}{!}{
\begin{NiceTabular}{cc}
\toprule 
OPE & exchanged representations with spin parity
\\
\midrule
$\phi \times \phi$   & $S_++ X_+ +\phi_+ +A_- + Z_++B_-$
\\
$X \times X$     & $S_+ +X_+ + (\overline{XX})_-$
\\
$Z_3 \times Z_3$    & $S_+ + (AA)_+ + (VAA)_+ + (AZ)_- + A_- + \phi_+ + (VA)_- + Z_+ + (Z_3)_+ + (AAA)_-$
\\
$\phi \times X$    & $\phi_\pm+(XV)_\pm$
\\
$\phi \times Z_3$    &  $Z_\pm+(AZ)_\pm+(VB)_\pm+(Z_3)_\pm$ 
\\
\bottomrule
\end{NiceTabular}
}
\end{table}

\begin{table}[t]
    \centering
    \caption{Gap assumptions imposed throughout most of the present work. Operators not mentioned below satisfy the unitarity bound ($d-2+\ell$ for spin $\ell$ primaries) plus a small twist gap $\tau = 10^{-6}$, i.e. $\Delta_{O} \geq \ell+\tau$. For singlets, we impose just unitarity since quasi-primaries that are Virasoro descendants of the identity have zero twist. Primes on operators denote subleading operators. Subscripts denote the spin. Motivation for these gaps is given in parts of the main text. {\bf When an assumption for an operator is given both in this table and the caption of a figure, then the assumption in the figure replaces the one of the table.}}\label{tab:gaps}
    \renewcommand{\arraystretch}{0.8}
    \begin{tabular}{cc}
      \toprule
      Operator & Gap Assumption\\ \midrule
      $S_0$   & $\geq 1$\\
      $T_{\mu\nu} = S_2$    & $=2$\\
      $S_2^\prime$    & $\geq 2.5$\\
      $S_{\ell = 1 \text{ or }\geq 3}$ & $\geq \ell$\\
      $X_0$    & $\geq 0.5$\\
      $\phi'_{0} = V^\prime_0$    &$\geq 0.7$\\
      $Z_0$ & $\geq 0.2$\\
      \bottomrule
    \end{tabular}
    \hspace{1em}
    \begin{tabular}{cc}
      \toprule
      Operator & Gap Assumption\\ \midrule
      $A_1$    & $\geq 1.1$\\
      $(VA)_1$ &$\geq 1.2$\\
      $(AZ)_1$ &$\geq 1.2$\\
      $(AA)_0$ &$\geq 2.2$\\
      $(VAA)_0 $ &$\geq 2.3$\\
      $(AAA)_0$ &$\geq 3.3$\\
      $({Z_3^\prime})_{0}$ &$\geq 1$\\
      \bottomrule
    \end{tabular}
\end{table}

\subsection{\texorpdfstring{$\phi$}{phi} Single Correlator Bootstrap}
\label{singlecorrresults}
\begin{figure}[H]
  \centering
  \includegraphics[scale=\defaultplotscale]{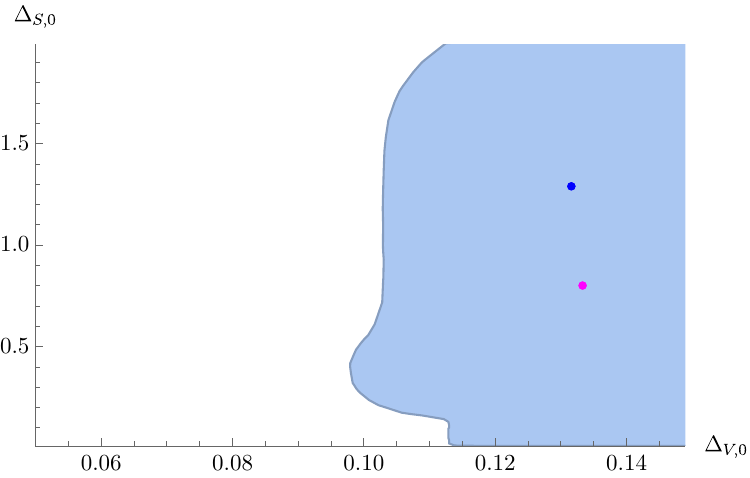}
  \caption{Single correlator ($\phi$ external) bound on the scaling dimension of the $O_{S,0}$ operator as a function of the scaling dimension of $\phi=O_{V,0}$. The numerical parameters used to obtain this plot are given in Appendix~\ref{numericalparameters}. The assumptions imposed are those of Table~\ref{tab:gaps} in addition to $\Delta_{S^\prime,0}\geq 2$. For more details see the discussion around and above Table~\ref{tab:gaps}.} \label{fig:Sspin0singlecorr}
\end{figure}

\begin{figure}[H]
  \centering
  \includegraphics[scale=\defaultplotscale]{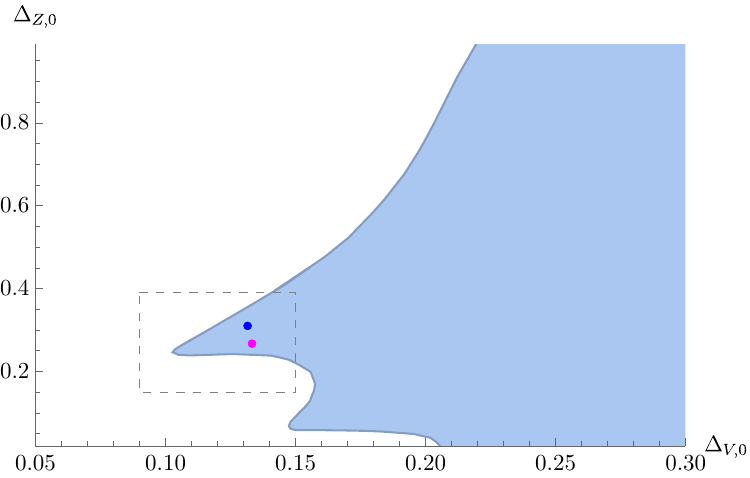}
  \caption{Single correlator ($\phi$ external) allowed region for the scaling dimension of the $O_{Z,0}$ operator as a function of the scaling dimension of $\phi=O_{V,0}$. The numerical parameters used to obtain this plot are given in Appendix~\ref{numericalparameters}. The assumptions imposed are those of Table~\ref{tab:gaps} in addition to $\Delta_{Z^\prime, 0}\geq 1$. For more details see the discussion around and above Table~\ref{tab:gaps}.} \label{fig:Zspin0singlecorr}
\end{figure}

\begin{figure}[H]
  \centering
  \includegraphics[scale=\defaultplotscale]{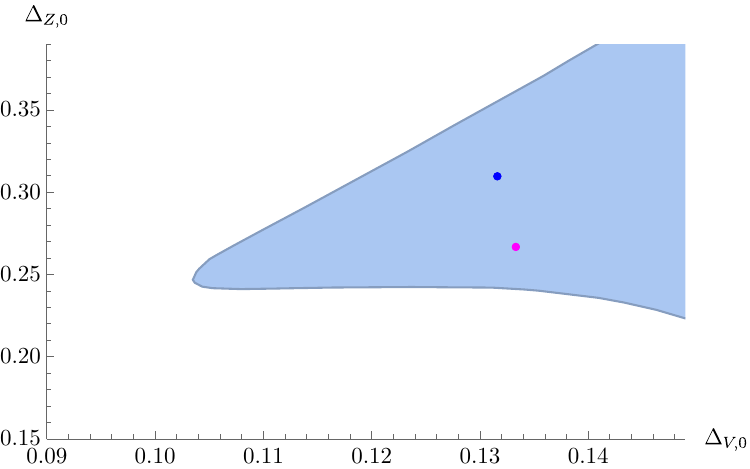}
  \caption{Zoomed in version of the peninsula in Figure~\ref{fig:Zspin0singlecorr}.}\label{fig:Zspin0singlecorrpeninsula}
\end{figure}

\begin{figure}[H]
  \centering
  \includegraphics[scale=\defaultplotscale]{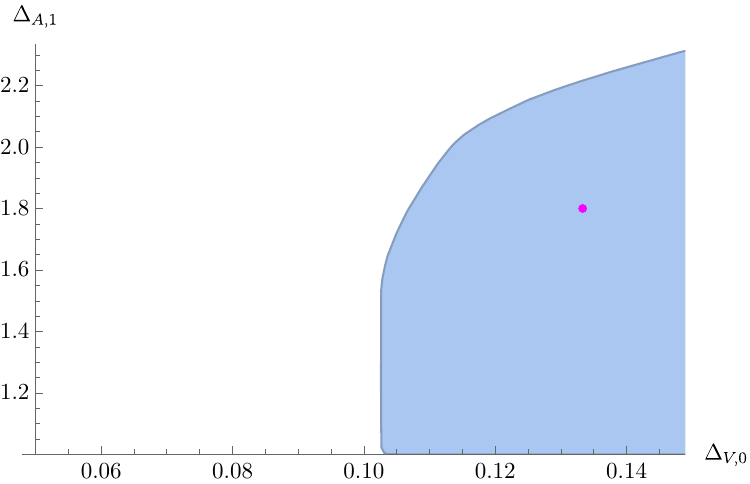}
  \caption{Single correlator ($\phi$ external) bound on the scaling dimension of the leading $O_{A,1}$ operator as a function of the scaling dimension of $\phi=O_{V,0}$. The numerical parameters used to obtain this plot are given in Appendix~\ref{numericalparameters}. The assumptions imposed are those of Table~\ref{tab:gaps}. For more details see the discussion around and above Table~\ref{tab:gaps}.} \label{fig:Aspin1singlecorr}
\end{figure}

\begin{figure}[H]
  \centering
  \includegraphics[scale=\defaultplotscale]{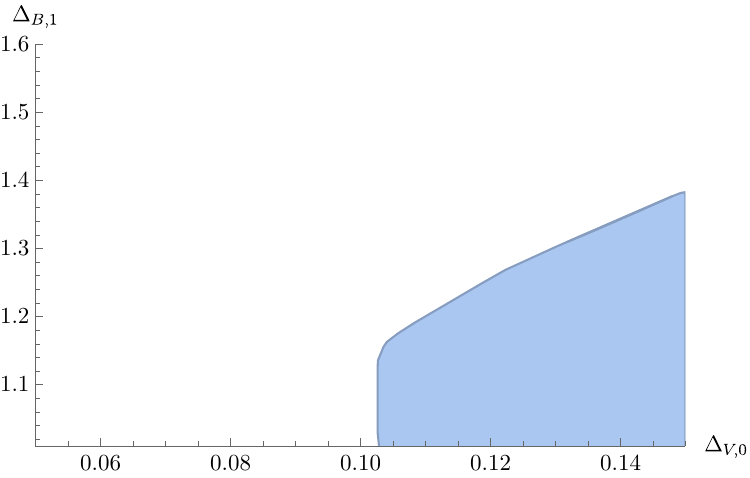}
  \caption{Single correlator ($\phi$ external) bound on the scaling dimension of the leading $O_{B,1}$ operator as a function of the scaling dimension of $\phi=O_{V,0}$. The numerical parameters used to obtain this plot are given in Appendix~\ref{numericalparameters}. The assumptions imposed are those of Table~\ref{tab:gaps}. For more details see the discussion around and above Table~\ref{tab:gaps}.} \label{fig:Bspin1singlecorr}
\end{figure}

\begin{figure}[H]
  \centering
  \includegraphics[scale=\defaultplotscale]{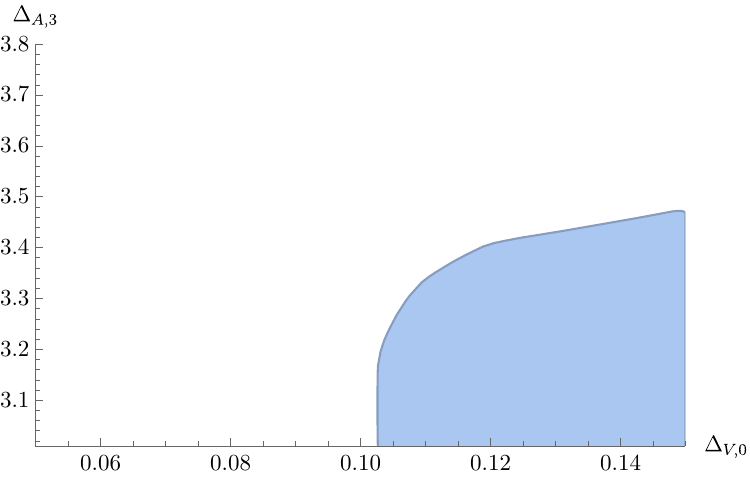}
  \caption{Single correlator ($\phi$ external) bound on the scaling dimension of the leading $O_{A,3}$ operator as a function of the scaling dimension of $\phi=O_{V,0}$. The numerical parameters used to obtain this plot are given in Appendix~\ref{numericalparameters}. The assumptions imposed are those of Table~\ref{tab:gaps}. We remind that the spin-$3$ operator $O_{A,3}$ is conserved at the decoupled fixed point. For more details see the discussion around and above Table~\ref{tab:gaps}.} \label{fig:Aspin3singlecorr}
\end{figure}

\begin{figure}[H]
  \centering
  \includegraphics[scale=\defaultplotscale]{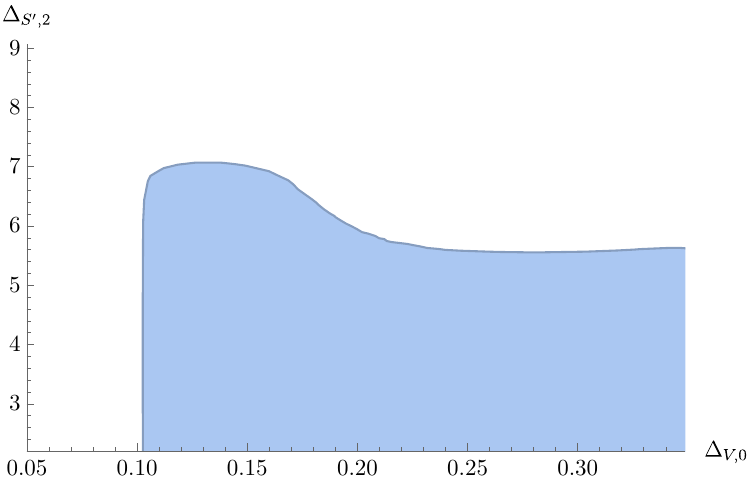}
  \caption{Single correlator ($\phi$ external) bound on the scaling dimension of the  $O_{S^\prime,2}$ operator, i.e.\ the subleading spin-$2$ operator after the stress-tensor, as a function of the scaling dimension of $\phi=O_{V,0}$. The numerical parameters used to obtain this plot are given in Appendix~\ref{numericalparameters}. The assumptions imposed are those of Table~\ref{tab:gaps}. For more details see the discussion around and above Table~\ref{tab:gaps}.} \label{fig:TmnPrimesinglecorr}
\end{figure}

\begin{figure}[H]
  \centering
  \includegraphics[scale=\defaultplotscale]{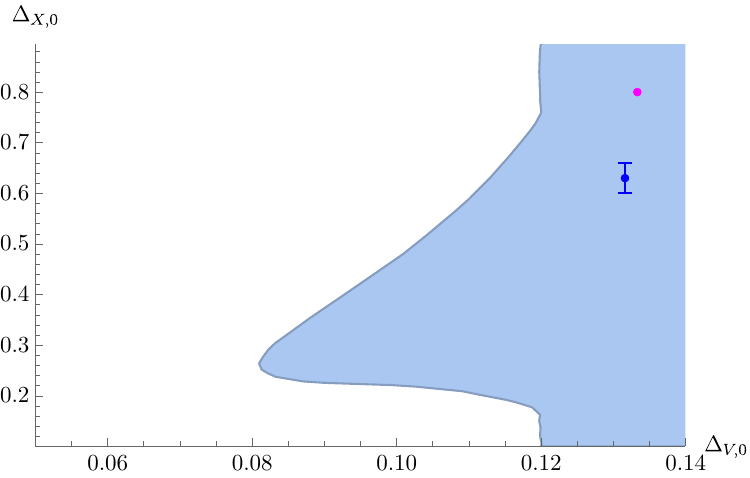}
  \caption{Single correlator ($\phi$ external) allowed region for the scaling dimension of the $O_{X,0}$ operator as a function of the scaling dimension of $\phi=O_{V,0}$. The numerical parameters used to obtain this plot are given in Appendix~\ref{numericalparameters}. The assumptions imposed are those of Table~\ref{tab:gaps} in addition to $\Delta_{X^\prime,0}\geq 0.75$. For more details see the discussion around and above Table~\ref{tab:gaps}.} \label{fig:Xspin0singlecorr}
\end{figure}

\begin{figure}[H]
  \centering
  \includegraphics[scale=\defaultplotscale]{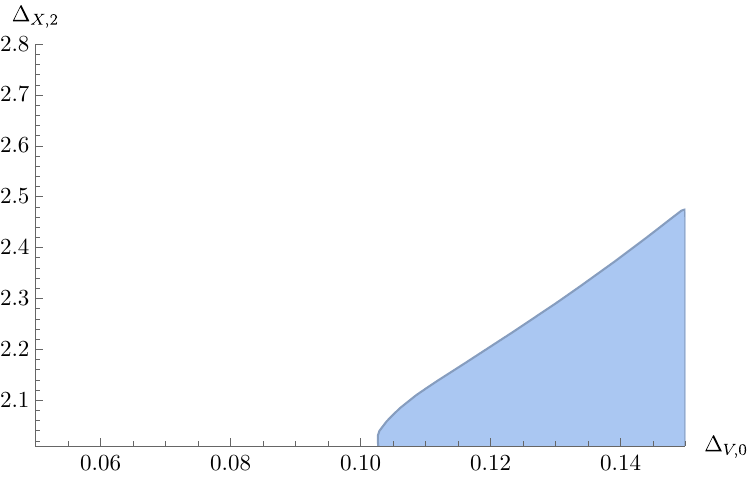}
  \caption{Single correlator ($\phi$ external) bound on the scaling dimension of the $O_{X,2}$ operator as a function of the scaling dimension of $\phi=O_{V,0}$. In the decoupled theory $O_{X,2}$ is a linear combination of the stress-tensors of the Potts models and is thus conserved. The numerical parameters used to obtain this plot are given in Appendix~\ref{numericalparameters}. The assumptions imposed are those of Table~\ref{tab:gaps}. For more details see the discussion around and above Table~\ref{tab:gaps}.} \label{fig:Xspin2singlecorr}
\end{figure}

\begin{figure}[H]
  \centering
  \includegraphics[scale=\defaultplotscale]{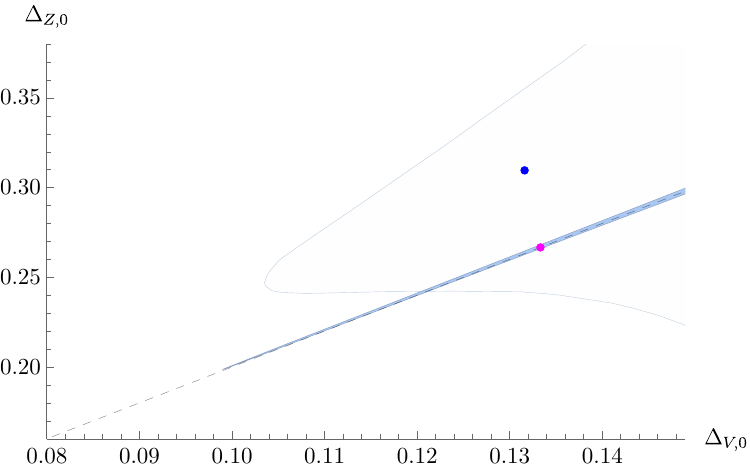}
  \caption{Single correlator ($\phi$ external) allowed region for the scaling dimension of the leading $O_{Z,0}$ operator as a function of the scaling dimension of $\phi=O_{V,0}$, both for $n=3$ and $n=100$, the second of which presenting the narrow daggerlike bound. The numerical parameters used to obtain this plot are given in Appendix~\ref{numericalparameters}. The assumptions imposed are those of Table~\ref{tab:gaps} in addition to $\Delta_{Z^\prime, 0}\geq 1$. For more details see the discussion around and above Table~\ref{tab:gaps}.} \label{fig:Zspin0singlecorrlargeN}
\end{figure}

\begin{figure}[H]
  \centering
  \includegraphics[scale=\defaultplotscale]{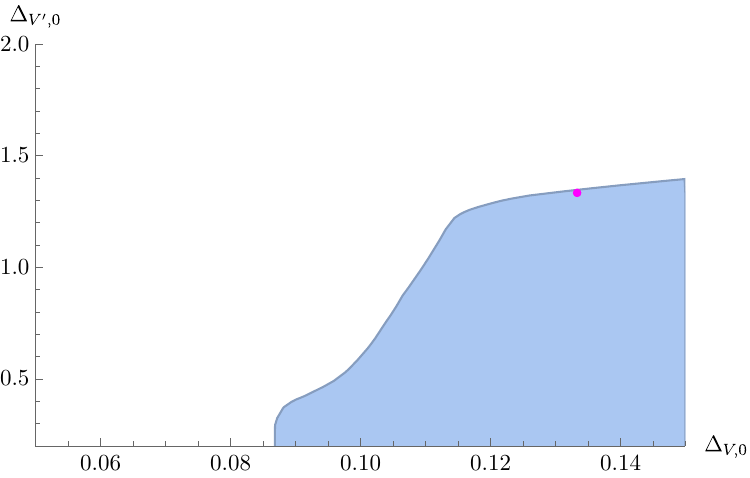}
  \caption{Single correlator ($\phi$ external) bound on the scaling dimension of the $O_{V^\prime,0}$ operator, i.e.\ the subleading operator after $O_{V,0}$ in the V spin-$0$ irrep, as a function of the scaling dimension of $\phi=O_{V,0}$. The numerical parameters used to obtain this plot are given in Appendix~\ref{numericalparameters}. The assumptions imposed are those of Table~\ref{tab:gaps}. For more details see the discussion around and above Table~\ref{tab:gaps}.}\label{fig:Vspin0singlecorr}
\end{figure}

In Figures~\ref{fig:Sspin0singlecorr} through~\ref{fig:Vspin0singlecorr} we show bounds obtained on some of the first few operators in the theory. This is done using the crossing equation in Eq.~\eqref{singlecrossing}. We use the notation $O_{R,\ell}$ where $R$ is the global symmetry irrep and $\ell$ the $SO(2)$ spin-label. The operators bounded are $O_{S,0}$, $O_{Z,0}$, $O_{A,1}$, $O_{B,1}$, $O_{A,3}$, $O_{S^\prime,2}$, $O_{X,0}$ $O_{X,2}$ and $O_{V^\prime, 0}$, whose scaling dimensions are constrained as a function of $\Delta_{\phi} = \Delta_{V,0}$. 

An important comment is that in all plots we have enforced several assumptions on operators, beyond the ones being plotted. Typically, one would like to start by obtaining bounds with no assumptions whatsoever (except for the operator dimension being bounded), and then progressively impose more assumptions in order to single out a theory of interest. We have found this not possible in our case because, in the absence of any assumptions, the bounds are trivial, i.e.\ there is no boundary between allowed and disallowed points, only allowed points. Hence the situation in $2d$ is different from the one typically encountered in $3d$. This phenomenon has already been observed in $2d$ in earlier work, for instance in the lower bound on the central charge \cite{Vichi:2011zza}. 

The assumptions we use are outlined in Table~\ref{tab:gaps} and in the captions of the figures. See the caption of each figure for the precise assumptions in each case. The gaps on $\Delta_{S,0}$, $\Delta_{X,0}$ and $\Delta_{Z,0}$ found in Table~\ref{tab:gaps} are motivated by transfer matrix results reported in Table~\ref{tab:epsexp}. The rest of the assumptions are motivated by perturbative intuition, i.e.\ perturbation theory around decoupled theories or the large-$n$ limit, and are somewhat conservative. For example, the first two sub-leading spin-$0$ $V$ operators are built with $\sigma \phi^a_i$ and $\phi^{\prime a}_i$, which both have a dimension of $4/3$ at zeroth order in the large-$n$ limit. The first is just the field $\phi^a_i$ multiplied by the auxiliary ``Hubbard-Stratonovich'' field, whereas the second is the subleading vector representation quasi-primary of the (single/decoupled) critical 3-state Potts model (see Appendix~\ref{AppQuasiPrimaries}). We generically expect the scaling eigenoperators to be some combination of these two building blocks, which will obtain an anomalous dimension.

A few comments are in hand. First, we observe that the bound in Figure~\ref{fig:Sspin0singlecorr} on the dimension of $O_{S,0}$ is rather unconstraining. We attribute this, at least in part, to the fact that this operator is relatively heavy. Secondly, and perhaps unsurprisingly, we observe that the most constraining bound is the one on the $O_{Z,0}$ operator as a function of $O_{V,0}$, Figure~\ref{fig:Zspin0singlecorr} and Figure~\ref{fig:Zspin0singlecorrpeninsula}. This is because these two operators are the two lightest in the theory, and the bootstrap is known to be the most constraining as close as possible to the unitarity bound ($(d-2)/2 = 0$ for scalars in $d=2$). In addition to the interacting theory, the bound can also be seen to include the decoupled theory, which we remind the reader has the same symmetry. Thus, unfortunately, while we seem to obtain a peninsula including our theory of interest we are unable to segregate it from other theories in the same parameter space, e.g.\ the decoupled theory. This is despite the gap imposed on the first singlet $\Delta_{S,0} \geq 1$ which would in principle exclude the decoupled theory, having the lightest singlet operator at $\Delta_{S,0} = 0.8$. We will improve on this later on in the present text by mixing in additional external operators. 

The rest of the bounds do not seem to present any striking features close to the expected fixed point. A sample of them includes the $O_{V,0}$-$O_{A,1}$ bound Figure~\ref{fig:Aspin1singlecorr} which includes the decoupled theory (and presumably the coupled one) in the bulk of the allowed region. The  $O_{V,0}$-$O_{B,1}$ bound Figure~\ref{fig:Bspin1singlecorr}. The $O_{V,0}$-$O_{A,3}$ bound Figure~\ref{fig:Aspin3singlecorr}, which is of interest since at the decoupled fixed point the $O_{A,3}$ operator becomes the spin-$3$ conserved current of the critical $3$-state Potts model. The $O_{V,0}$-$O_{S^\prime,2}$ bound Figure~\ref{fig:TmnPrimesinglecorr}, which bounds the sub-leading operator after the stress tensor. While this plot has a significant feature around the value of $O_{V,0}$ corresponding to our CFT of interest, we do not attribute it to our theory of interest given the values of $\Delta_{S^\prime,2}$ which it is located at. Consequently, we observe that while the $O_{V,0}$-$O_{X,0}$ bound Figure~\ref{fig:Xspin0singlecorr} has a feature, it does not seem to be close to our theory of interest. Indeed, we will see it disappears later when using $\phi - X$ mixed correlator system Eq.~\eqref{crossingPhiX} (Figures~\ref{fig:Xspin0PhiX} and~\ref{fig:Xspin0PhiX11vs19}). We also show the $O_{V,0}$-$O_{X,2}$ bound Figure~\ref{fig:Xspin2singlecorr}, where $O_{X,2}$ becomes conserved at the decoupled fixed point,\footnote{Since it is a linear combination of stress tensors in the individual Potts replicas.} but is expected to obtain an anomalous dimension at the interacting one.

In Figure~\ref{fig:Zspin0singlecorrlargeN} we overlay the bounds on $\Delta_{Z,0}$ obtained at $n=3$ and at $n=100$ (all previous and subsequent plots concern $n=3$). As $n \rightarrow \infty $, $\Delta_{Z,0}$ should approach $2\Delta_\phi$. Indeed, we observe that the peninsula shrinks to a very thin ``dagger'', which we relate to the fact that the coupled and decoupled theories sit on top of each other in this limit, and in this particular slice of parameter space. The slope of the dagger is $\Delta_{Z,0}=2\Delta_{\phi}$, as expected. This reinforces the large-$n$ intuition for the theory. However, as also discussed earlier, we have no knowledge of the $1/n$ corrections, and as indicated by transfer matrices and our plots later on, certain operators do acquire rather large anomalous dimensions. For this reason, we opt to use assumptions that are somewhat conservative, such that our plots present somewhat ``universal'' bounds that are not too heavily gap-dependent. Lastly, in Figure~\ref{fig:Vspin0singlecorr} we present the $O_{V,0}$-$O_{V^\prime, 0}$ exclusion bound. This is saturated by the decoupled theory, similarly to the corresponding plot in~\citep{Rong:2017cow} studying CFTs with $S_3$ symmetry.

\subsection{\texorpdfstring{$Z_3$}{Z3} Single Correlator Bootstrap}
\begin{figure}[H]
  \centering
  \includegraphics[scale=\defaultplotscale]{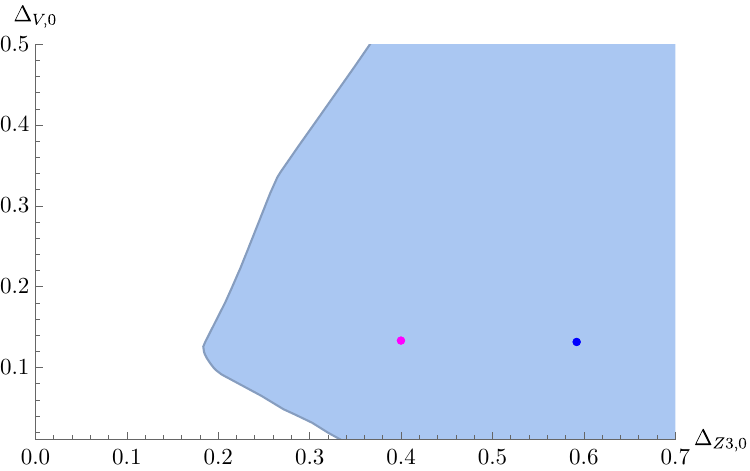}
  \caption{Single correlator ($Z_3$ external) allowed region for the scaling dimension of $O_{V,0}$ operator as a function of the scaling dimension of $Z_3= O_{Z_3,0}$. The numerical parameters used to obtain this plot are given in Appendix~\ref{numericalparameters}. The assumptions imposed are those of Table~\ref{tab:gaps}. For more details see the discussion around and above Table~\ref{tab:gaps}.} \label{fig:Vspin0Z3singlecorr}
\end{figure}

\begin{figure}[H]
  \centering
  \includegraphics[scale=\defaultplotscale]{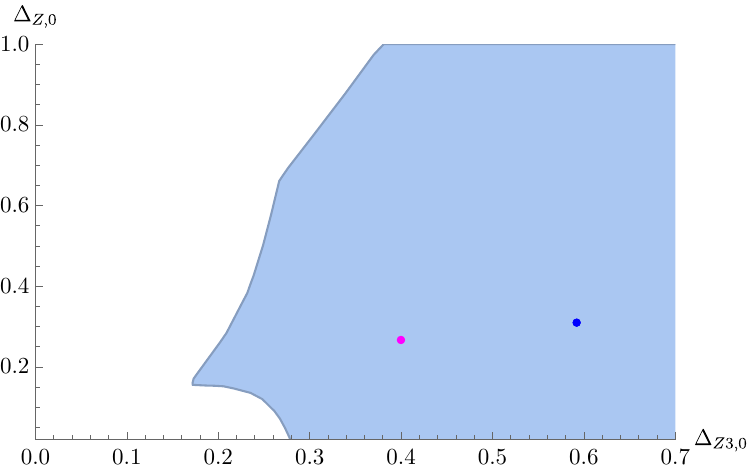}
  \caption{Single correlator ($Z_3$ external) allowed region for the scaling dimension of $O_{Z,0}$ operator as a function of the scaling dimension of $Z_3 = O_{Z_3,0}$. The numerical parameters used to obtain this plot are given in Appendix~\ref{numericalparameters}. The assumptions imposed are those of Table~\ref{tab:gaps}. For more details see the discussion around and above Table~\ref{tab:gaps}.} \label{fig:Zspin0Z3singlecorr}
\end{figure}

We briefly showcase in Figure~\ref{fig:Vspin0Z3singlecorr} the bound on $\Delta_{V,0}$ obtained using the $Z_3$ single correlator system in Eq.~\eqref{crossingZ3single}. This will serve to demonstrate the improvement achieved when mixing both $\phi$ and $Z_3$ later on using the mixed correlator system in Eq.~\eqref{crossingPhiZ3}. Similarly we can also bound $\Delta_{Z,0}$ as a function of $\Delta_{Z_3,0}$ using the same correlator system, which we showcase in Figure \ref{fig:Zspin0Z3singlecorr}.

In addition to assumptions outlined in Subsection~\ref{singlecorrresults}, we have imposed some new ones due to the new irreps that can appear in the OPE involving $Z_{3}$. These are reported again in Table~\ref{tab:gaps}. An example is $\Delta_{AAA,0} \geq 3.3$. To motivate these assumptions it helps to think of the ``composite'' irreps discussed in Subsection~\ref{compositeirreps}. At the decoupled fixed point (or large-$n$), the operator $O_{AAA,0}$ must be built by taking an $A$ operator from each of the three replicas and symmetrizing them (possibly with insertions of derivatives). The $A$ operators of each replica can be of any spin as long as the combination reduces to spin-$0$. Thus the lowest lying $O_{AAA,0}$ operator at the decoupled (or large-$n$) fixed point cannot be lighter than the combination of the three lightest $A$ operators, i.e.\ $3\times 1.8$,\footnote{For the quasi-primaries see Appendix~\ref{AppQuasiPrimaries}.} which is much bigger than the gap $\Delta_{AAA,0} \geq 3.3$ we impose.

\subsection{\texorpdfstring{$\phi -X$}{phi-X} Mixed Correlator Bootstrap}
\begin{figure}[H]
  \centering
  \includegraphics[scale=\defaultplotscale]{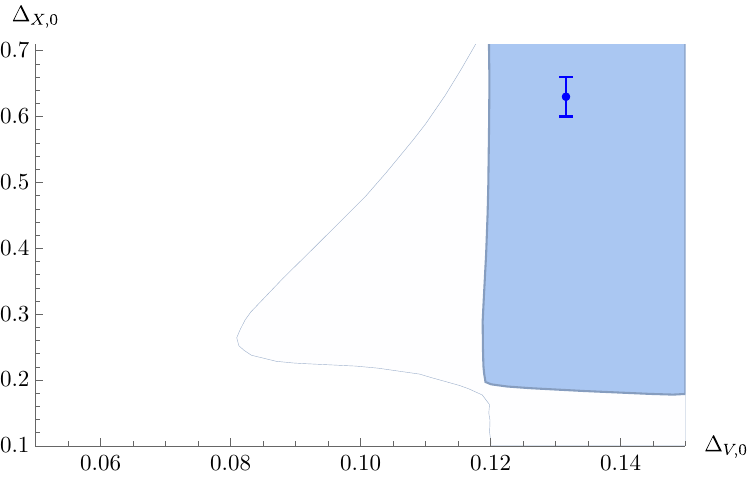}
  \caption{Single versus mixed correlator ($\phi - X$ externals) allowed region for scaling dimensions of ${X=O_{X,0}}$ and $\phi=O_{V,0}$. The numerical parameters used to obtain this plot are given in Appendix~\ref{numericalparameters}. The assumptions imposed are those of Table~\ref{tab:gaps} in addition to $\Delta_{X^\prime,0}\geq 0.75$. For more details see the discussion around and above Table~\ref{tab:gaps}. We also fix the ratio of OPE coefficients ${\lambda_{\phi \phi T_{\mu \nu}}/\lambda_{X X T_{\mu \nu}}=\Delta_{\phi} / \Delta_{X}}$.} \label{fig:Xspin0PhiX}
\end{figure}

\begin{figure}[H]
  \centering
  \includegraphics[scale=\defaultplotscale]{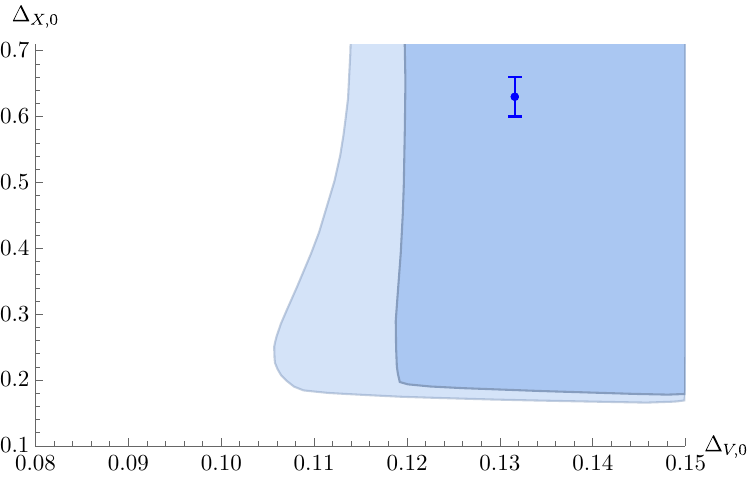}
  \caption{Mixed correlator ($\phi - X$ externals) allowed region for scaling dimensions of ${X=O_{X,0}}$ and ${\phi=O_{V,0}}$. The numerical parameters used to obtain this plot are given in Appendix~\ref{numericalparameters}, in particular, we obtain the plot at two different values of $\Lambda$ ($\Lambda=11$ and $19$) to illustrate convergence. Shaded areas present the admissible parameter space, surrounded by the excluded region. The assumptions imposed are those of Table~\ref{tab:gaps} in addition to $\Delta_{X^\prime,0}\geq 0.75$. For more details see the discussion around and above Table~\ref{tab:gaps}. We also fix the ratio of OPE coefficients ${\lambda_{\phi \phi T_{\mu \nu}}/\lambda_{X X T_{\mu \nu}}=\Delta_{\phi} / \Delta_{X}}$.} \label{fig:Xspin0PhiX11vs19}
\end{figure}

\begin{figure}[H]
  \centering
  \includegraphics[scale=\defaultplotscale]{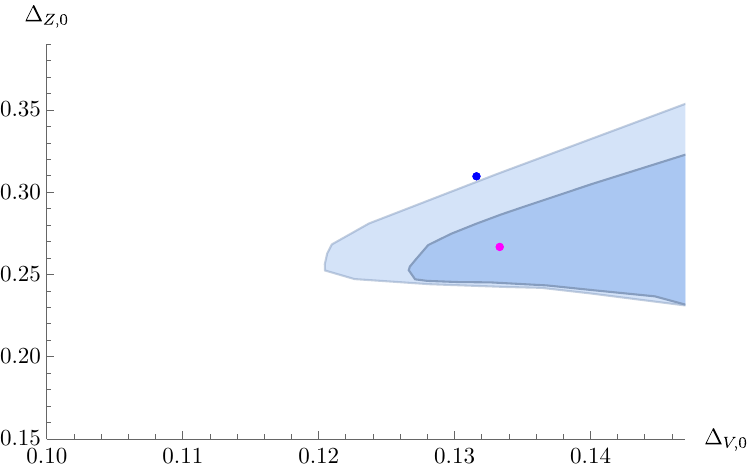}
  \caption{Mixed correlator ($\phi - X$ externals) allowed region for the scaling dimension of the leading $O_{Z,0}$ operator as a function of the scaling dimension of $\phi=O_{V,0}$. In order to obtain the plot we projected the $\Delta_{V,0}$-$\Delta_{X,0}$-$\Delta_{Z,0}$ allowed space onto the $\Delta_{V,0}$-$\Delta_{Z,0}$ subspace. The numerical parameters used to obtain this plot are given in Appendix~\ref{numericalparameters}. The assumptions imposed are those of Table~\ref{tab:gaps} in addition to $\Delta_{Z^\prime, 0}\geq 1$ and $\Delta_{X^\prime,0}\geq 0.75$ or $\Delta_{X^\prime,0}\geq 1$. For more details see the discussion around and above Table~\ref{tab:gaps}. We also fix the ratio of OPE coefficients ${\lambda_{\phi \phi T_{\mu \nu}}/\lambda_{X X T_{\mu \nu}}=\Delta_{\phi} / \Delta_{X}}$.} \label{fig:Zspin0PhiX}
\end{figure}

Firstly, in Figure~\ref{fig:Xspin0PhiX} we present the mixed correlator $O_{V,0}$-$O_{X,0}$ exclusion plot. This is obtained by mixing in $O_{X,0}$ as an external, which corresponds to crossing Eq.~\eqref{crossingPhiX}. We superimpose the corresponding single correlator plot, Figure~\ref{fig:Xspin0singlecorr}, to illustrate that the apparent feature has gone away. In Figure~\ref{fig:Xspin0PhiX11vs19}, we present the same plot but at two different values of $\Lambda$, $\Lambda=11$ and $19$. The apparent feature is again observed to disappear. We now proceed by taking a closer look at the $O_{V,0}$-$O_{Z,0}$ allowed region in Figure~\ref{fig:Zspin0PhiX}, which is the projection of the allowed region in $\Delta_{V,0}$-$\Delta_{X,0}$-$\Delta_{Z,0}$ space onto the $\Delta_{V,0}$-$\Delta_{Z,0}$ plane. The projection is done for the slice $0.55 < \Delta_{X,0} < 0.75$ which corresponds to the area around the coupled replica CFT, see Table~\ref{tab:epsexp}. The two bounds in Figure~\ref{fig:Zspin0PhiX} differ in the gap assumption on $O_{X^\prime, 0}$, which is the gap to which we found the plots to be the most sensitive. In particular, we find that the blue dot, corresponding to the interacting fixed point, seems to be excluded both for $\Delta_{X^\prime, 0} \geq 0.75$ (slightly) and for  $\Delta_{X^\prime, 0} \geq 1$. We recall that at the decoupled fixed $\Delta_{X^\prime,0}=1.6$. It would thus appear that $\Delta_{X^\prime, 0}$ acquires a large negative anomalous dimension. The operator does  acquire a negative anomalous dimension in perturbation theory, see Table~\ref{tab:epsexp}. However, since the results are known only to two orders in $\varepsilon$ it is hard to draw a definitive conclusion. Notice also that the coefficient of the $\varepsilon^2$ term has a large negative value, which may also point in this direction.

\subsection{\texorpdfstring{$\phi - Z_3$}{phi-Z3} Mixed Correlator Bootstrap}
Finally, we proceed to the constraints obtained using the crossing system in Eq.~\eqref{crossingPhiZ3}. As we will see Eq.~\eqref{crossingPhiZ3} carries the advantage of being able to differentiate between the coupled and decoupled fixed points. In other words, we can obtain an allowed region that contains the coupled fixed point while the decoupled one being disallowed. This is the main advantage of this system over Eq.~\eqref{crossingPhiX}. The improvement is perhaps in part due to the operator $O_{Z_3,0}$ gaining a relatively large anomalous dimension. Notice that this anomalous dimension is especially large compared to the one $O_{Z,0}$ obtains, see Table~\ref{tab:epsexp}.\footnote{For the reader's convenience we remind that these operators have dimensions $2\Delta_{\phi}$ and $3\Delta_{\phi}$ at the decoupled fixed point.} 

\begin{figure}[H]
  \centering
  \includegraphics[scale=\defaultplotscale]{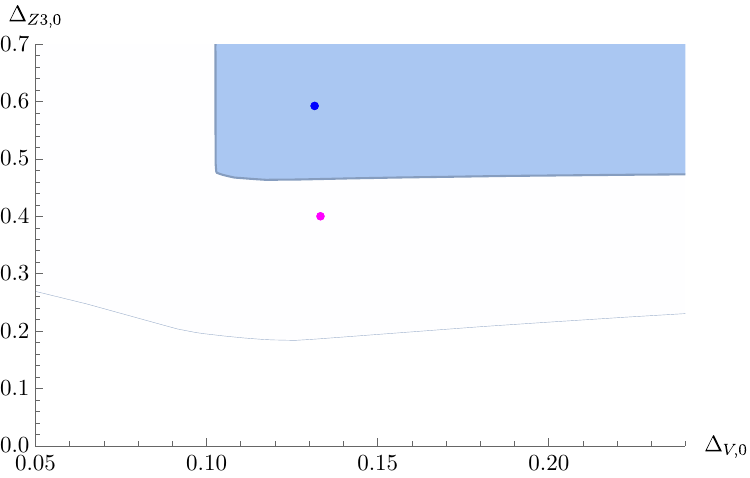}
  \caption{Mixed correlator ($\phi - Z_{3}$ externals) allowed region for scaling dimensions of $Z_3=O_{Z_3,0}$ and ${\phi = O_{V,0}}$. The numerical parameters used to obtain this plot are given in Appendix~\ref{numericalparameters}. The assumptions imposed are those of Table~\ref{tab:gaps}. For more details see the discussion around and above Table~\ref{tab:gaps}. We also fix the ratio of OPE coefficients ${\lambda_{\phi \phi T_{\mu \nu}}/\lambda_{Z_3 Z_3 T_{\mu \nu}}=\Delta_{\phi} / \Delta_{Z_3}}$.} \label{fig:Z3spin0FromPhiZ3}
\end{figure}

\begin{figure}[H]
  \centering
  \includegraphics[scale=\defaultplotscale]{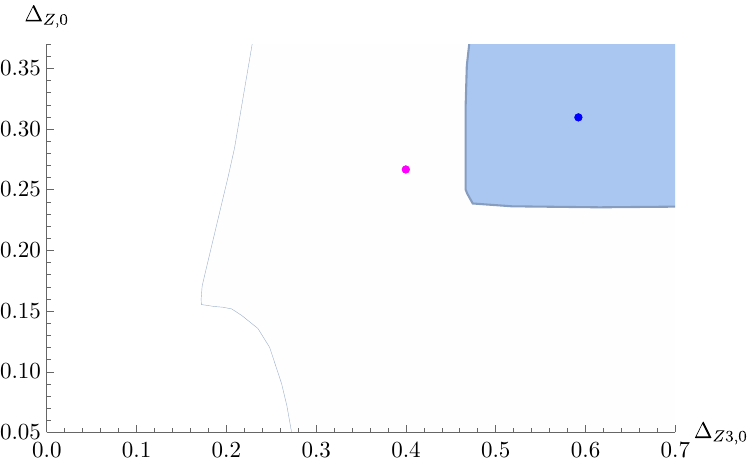}
  \caption{Mixed correlator ($\phi - Z_{3}$ externals) allowed region for $O_{Z,0}$ operator as a function of the scaling dimension of $Z_3=O_{Z_3,0}$. To obtain the plot we projected the $\Delta_{V,0}$-$\Delta_{Z,0}$-$\Delta_{Z_3,0}$ allowed space onto the $\Delta_{Z_3,0}$-$\Delta_{Z,0}$ subspace. The numerical parameters used to obtain this plot are given in Appendix~\ref{numericalparameters}. The assumptions imposed are those of Table~\ref{tab:gaps} in addition to $\Delta_{Z^\prime, 0}\geq 1$. For more details see the discussion around and above Table~\ref{tab:gaps}. We also fix the ratio of OPE coefficients ${\lambda_{\phi \phi T_{\mu \nu}}/\lambda_{Z_3 Z_3 T_{\mu \nu}}=\Delta_{\phi} / \Delta_{Z_3}}$.} \label{fig:Z3vsZspin0FromPhiZ3}
\end{figure}

\begin{figure}[H]
  \centering
  \includegraphics[scale=\defaultplotscale]{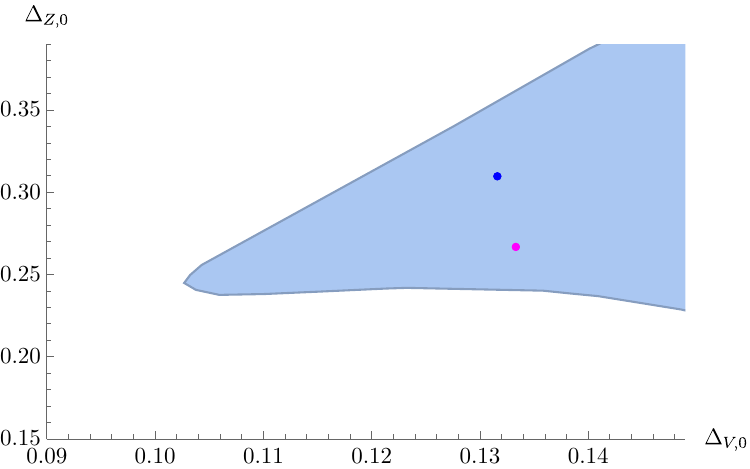}
  \caption{Mixed correlator ($\phi - Z_{3}$ externals) allowed region for $O_{Z,0}$ operator as a function of the scaling dimension of $\phi=O_{V,0}$. In order to obtain the plot we projected the $\Delta_{V,0}$-$\Delta_{Z,0}$-$\Delta_{Z_3,0}$ allowed space onto the $\Delta_{V,0}$-$\Delta_{Z,0}$ subspace. The numerical parameters used to obtain this plot are given in Appendix~\ref{numericalparameters}. The assumptions imposed are those of Table~\ref{tab:gaps} in addition to $\Delta_{Z^\prime, 0}\geq 1$. For more details see the discussion around and above Table~\ref{tab:gaps}. We also fix the ratio of OPE coefficients ${\lambda_{\phi \phi T_{\mu \nu}}/\lambda_{Z_3 Z_3 T_{\mu \nu}}=\Delta_{\phi} / \Delta_{Z_3}}$.} \label{fig:Zspin0FromPhiZ3}
\end{figure}

In Figure~\ref{fig:Z3spin0FromPhiZ3} we show the $\Delta_{V,0}$-$\Delta_{Z_3,0}$ allowed region. On the other hand, in Figures~\ref{fig:Z3vsZspin0FromPhiZ3} and~\ref{fig:Zspin0FromPhiZ3}, we present the projections of the $\Delta_{V,0}$-$\Delta_{Z,0}$-$\Delta_{Z_3,0}$ allowed region onto two different planes. In Figure~\ref{fig:Z3spin0FromPhiZ3} ($\Delta_{V,0}$-$\Delta_{Z_3,0}$ plane) and in  Figure~\ref{fig:Z3vsZspin0FromPhiZ3} ($\Delta_{Z_3,0}$-$\Delta_{Z,0}$ plane) we explicitly observe that the decoupled theory (magenta dot) is excluded, while the coupled theory (blue dot) is included in the allowed region. Thus, while in the projection of Figure~\ref{fig:Zspin0FromPhiZ3} ($\Delta_{V,0}$-$\Delta_{Z,0}$ plane) it would naively appear that the decoupled point is allowed, we find out that it is not. We have thus found a sub-region of parameter space where the interacting theory is present and the decoupled is not.

While the $\Delta_{V,0}$-$\Delta_{Z,0}$ plot obtained via the projection of $\Delta_{V,0}$-$\Delta_{Z,0}$-$\Delta_{Z_3,0}$ does not show any particular improvement versus the corresponding plot obtained from the single correlator system Eq.~\eqref{singlecrossing}, the $\Delta_{Z_3,0}$-$\Delta_{Z,0}$ projection compared to the plot from the single correlator system Eq.~\eqref{crossingZ3single} does. Additionally, the $\Delta_{V,0}$-$\Delta_{Z_3,0}$ exclusion bound can either be obtained directly by placing a gap $\Delta_{Z,0}\geq 0.2$, or by adding $O_{Z,0}$ as an isolated exchanged operator and imposing $\Delta_{Z',0} \geq 1$ (and then projecting). We have found no noticeable difference between the two methods. We emphasise that Figure~\ref{fig:Z3spin0FromPhiZ3} is the first of the two options, i.e.\ we just impose $\Delta_{Z,0}\geq 0.2$.

{\bf A note on the navigator~\citep{Reehorst:2021ykw} and skydiving~\citep{Liu:2023elz} methods.} While we did experiment with using the navigator and skydiving methods, specifically in an analogue of Figure~\ref{fig:Zspin0FromPhiZ3}, where we also scanned over the OPE vectors ($\lambda_{VZ_3 Z_3}$, $\lambda_{VV V}$, $\lambda_{Z_3 Z_3 Z_3}$) and ($\lambda_{VVZ}$,$\lambda_{VZ_3 Z}$,$\lambda_{Z_3 Z_3 Z}$), we did not find any significant improvement both in terms of numerical efficiency as well as constraining power. In fact, the computations became considerably more expensive. Thus, we omit any discussion of these methods as applied to our problem. Let us stress, however, that we do not exclude the possibility that these methods could become more practical at higher values of $\Lambda$ (which controls the amount of components in the bootstrap functional). 

\section{Discussion and Outlook}
\label{sec:Discussion}
In the present work, we initiated a study of the space of admissible two-dimensional CFTs with ${S_n \ltimes (S_Q)^n}$ global symmetry, with a particularly strong emphasis on the $n=3, Q=3$ case. Our main results concern bounds on the space of allowed dimensions for the four lightest (in the UV)\footnote{These operators may as well prove to be the four lightest also in the IR, however without precise knowledge of the anomalous dimensions it is not certain.} operators in the theory. These results were derived by studying two mixed correlator systems, namely the $\phi - X$ and $\phi - Z_{3}$ systems. The first of which contains the 1\ap{st} and 4\ap{th} lightest operators in the theory and the second the 1\ap{st} and 3\ap{rd} lightest. From the $\phi - X$ systems, we seem to have (re)discovered that the second operator in the spin-$0$ $X$ representation, $X^\prime$, has a rather large negative anomalous dimension. This appears to be hinted at from perturbative results, see Table~\ref{tab:epsexp}. However, we are not currently aware of any non-perturbative predictions for comparison. Next, using the $\phi - Z_{3}$ system we were able to constrain the theory of interest within an allowed peninsula in parameter space, see Figure~\ref{fig:Zspin0FromPhiZ3}. Most importantly, we were able to exclude the decoupled $S_3 \ltimes (S_3)^3$ fixed point (Figures \ref{fig:Z3spin0FromPhiZ3} and \ref{fig:Z3vsZspin0FromPhiZ3}), which corresponds to three decoupled replicas of the critical $3$-state Potts model. This peninsula, however, did not show any signs of becoming an island when increasing the number of derivatives or using stronger assumptions.\footnote{While not included in this manuscript, we did try imposing some large gap assumptions inspired by the dimensions of quasi-primaries at the decoupled fixed point. See more generally the discussion in Appendix~\ref{AppQuasiPrimaries}.}

In practice, we found that our study was impeded by the lack of detailed knowledge on the spectrum of anomalous dimensions in the theory. With the exception of some earlier results (Table~\ref{tab:epsexp}) from perturbation theory and non-perturbative techniques, such as transfer matrices and Monte Carlo, not much is known about the precise spectrum. To a large extent, this is understandable, since for example operators in representations such as $AAA$  in Table~\ref{tab:irreps} are probably of no practical use outside the conformal bootstrap. However, in the numerical conformal bootstrap, precise knowledge of such operators can inform detailed gap assumptions on the spectrum which can lead to much more constraining results. We saw that in cases where we did have knowledge of the spectrum, specifically for the leading spin-$0$ $\phi$, $X$, $Z$ and $Z_3$ operators, we were able to make more progress. It goes without saying that further perturbative and non-perturbative results in the spirit of~\citep{Dotsenko:1998gyp} would be invaluable.

Another promising direction will be to study these theories through a quantum critical spin-chain in the spirit of~\citep{Milsted:2017csn}. In this approach, one may realise the Virasoro generators in terms of lattice degrees of freedom allowing one to single out both Virasoro primary and quasi-primary operators. 

Given that we are dealing with two-dimensional conformal field theories, another obvious direction is take advantage of the full conformal symmetry. That is, proceeding with a generalisation of the usual numerical conformal bootstrap technique using Virasoro conformal blocks, from which we expect more constraining results. This is because, under Virasoro, the spectrum of primaries in the theories is expected to be much more sparse. Moreover, since Virasoro blocks depend explicitly on the central charge $c$, this is a parameter that can then be scanned over. It would thus be interesting to be able to compare with the estimate of $c \simeq 2.36$ obtained in~\citep{Dotsenko:1998gyp} for 3 coupled 3-state Potts models. We hope that Virasoro blocks, together with more constraining assumptions inspired by results from independent methods, will provide the edge needed to obtain an island in parameter space. We aim to report on these issues in the near future.

Two more directions to consider would be to study the theories in question through a ``truncation method'', for example along the lines of~\citep{Kantor:2021jpz}, and also to study more complicated mixed correlators (such as mixing in as an external the leading spin-$0$ $Z$ operator). In the former case, one truncates the sum rule assuming a fixed number of exchanged operators and then tries to best determine their dimensions so that the sum rule is satisfied as best as possible, an idea going back to~\citep{Gliozzi:2013ysa}. This would provide data for anomalous dimensions needed when imposing gaps in the ``vanilla'' conformal bootstrap, either Global or Virasoro. On the point of more mixed correlators, e.g.\ including $Z$, these could prove useful since they can both provide more constraining power and also because different mixed correlators can exchange different representations from Table~\ref{tab:irreps}. This is something to consider because some representations can have (much) larger gaps than others, which would aid in obtaining an island in parameter space. Of course, the benefits of mixing in additional operators should be weighed against the increased numerical cost.

Lastly, let us note that our analysis has been carried out in a way that will allow us to recycle a lot of technology in the future when studying other theories. One such example are theories with $S_3 \ltimes (O(2))^3$ global symmetry, which are interesting due to applications in critical phenomena~\citep{DMukamel_1975}.

\acknowledgments
We thank Ning Su for numerous helpful conversations and help with simpleboot, we also thank Scott Collier for correspondence on unpublished results.
The research work of AV and SRK received funding from
the European Research Council (ERC) under the European Union's Horizon 2020 research
and innovation programme (grant agreement no. 758903). SRK would also like to acknowledge support from INFN Pisa. Computations for this work were run on the SISSA HPC cluster Ulysses and the INFN Pisa HPC cluster Theocluster Zefiro. AP and SRK benefited from attending the Mini-Course of Numerical Conformal Bootstrap\footnote{\href{https://perimeterinstitute.ca/events/mini-course-numerical-conformal-bootstrap}{https://perimeterinstitute.ca/events/mini-course-numerical-conformal-bootstrap}} hosted at Perimeter Institute. AP and SRK would also like to acknowledge that this research was supported in part by Perimeter Institute for Theoretical Physics. Research at Perimeter Institute is supported in part by the Government of Canada through the Department of Innovation, Science and Economic Development Canada and by the Province of Ontario through the Ministry of Colleges and Universities.

\appendix

\section{Calculation of Tensor Structures and Sum Rules}\label{apx:sumrules}
Let us note that in all the following we will use a particular choice of notation to simplify the presentation as much as possible. For example a sum rule of the form 
\label{sec:SumRules}
\begin{equation}
    0=\sum_{O} \left( \lambda_{ijO} \lambda_{klO} F^{ij,kl}_{\mp,\Delta,l}\pm \lambda_{kjO} \lambda_{ilO} F^{kj,il}_{\mp,\Delta,l}\right) \,,
    \label{crossingformula} 
\end{equation}
found in~\citep{Kos:2014bka}, which in the case of a single correlator Ising bootstrap would reduce to
\begin{equation}
    0=\sum_{O} \left( \lambda_{\phi \phi O} ^2 F^{\phi \phi,\phi \phi}_{-,\Delta,l}\right) \,,
\end{equation}
will be in this work shortened to 
\begin{equation}
    0=F^-_S.
\end{equation}
This will render larger expressions significantly more readable. In cases where there might be ambiguities in deducing what OPE coefficients should be multiplying the convolved conformal blocks $F$, we will state it explicitly. This can happen in mixed correlator settings, such as the crossing equation $\langle O O O^\prime O^\prime \rangle = \langle O^\prime O O O^\prime \rangle$, where if both OPEs $O \times O$ and $O \times O^\prime$ exchange an operator in common, then there would be an ambiguity to assigning OPE coefficients in our notation. We hope that this notation will not be confusing to the reader.

\subsection{Tensor Structures for \texorpdfstring{$\braket{\phi \phi \phi \phi}$}{<phi phi phi phi>}}
\label{singcorrphi}
The tensor structures for a correlator of four fields in the defining representation
\begin{equation}
\langle \phi^a_i \phi^b_j \phi^c_k \phi^d_l \rangle
\end{equation} 
can be easily worked out applying the rules of~\citep{Kousvos:2021rar}. In order to obtain the structures for $S_n \ltimes (S_Q)^n$, we start with those of $S_Q$. The OPE of two fundamentals $\phi_i$, $\phi_j$ of $S_Q$ decomposes as
\begin{equation}
\phi_i \times \phi_j \sim \left(\delta_{ij}-\frac{1}{Q}\omega_i \omega_j \right) S + C_{ijk} \phi_k+ T_{ij} + A_{ij} 
\end{equation}
for $Q \geq 4$, and
\begin{equation}
\phi_i \times \phi_j \sim \left(\delta_{ij}-\frac{1}{Q}\omega_i \omega_j \right) S  + c_{ijk} \phi_k + A_{ij}
\end{equation}
for $Q=3$. For $Q=2$ the OPE reduces to the one of an Ising-like CFT. Here $\omega_i$ is a column vector of units, $\sum_{i=1}^Q \omega_i=Q$. The irreps on the RHS include the singlet $S$, the defining irrep $\phi_k$ which is both external and exchanged, a two-index symmetric irrep $T_{ij}$ and a two-index antisymmetric irrep $A_{ij}$. OPEs of this type have been studied in~\citep{Hogervorst:2016itc,Rong:2017cow,Stergiou:2018gjj}. That being said, we will use a slightly different notation, including the vector $\omega_i$, to avoid a mismatch of indices. The tensor $c_{ijk}$ is proportional to $\langle \phi_i \phi_j \phi_k \rangle$ and satisfies $c_{ijr} c_{klr} = P^V_{ijkl}$ (to be defined below). The tensor structures, and projectors, for these irreps are 
\begin{equation}
\begin{aligned}
P^{S}_{ijkl} &= \frac{1}{Q-1}P_{ij} P_{kl} \,,\\
P^{V}_{ijkl} &= \frac{Q}{Q-2} d_{ijm} d_{klme}\,,\\
P^{T}_{ijkl} &= \frac{P_{ik}P_{jl}+P_{il}P_{jk}}{2}- \frac{Q}{Q-2} d_{ijm} d_{klm}-\frac{1}{Q-1} P_{ij} P_{kl} \,,\\
P^{A}_{ijkl} &= \frac{P_{ik} P_{jl} - P_{il} P_{jk}}{2} \,,\\
\end{aligned}
\label{singlecorrprojectors}
\end{equation}
where $d_{ijk}=\delta_{ijk}-\frac{1}{Q}\left(\delta_{ij}\omega_k +\delta_{ik}\omega_j +\delta_{kl}\omega_i \right)+\frac{2}{Q^2}\omega_i \omega_j \omega_k  $ and $P_{ij}=\delta_{ij}-\frac{1}{Q}\omega_i \omega_j$. Consequently, the projectors for the full group $S_n \ltimes (S_Q)^n$ are
\begin{equation}
\begin{aligned}
{P^S}^{\lsp abcd}_{\lsp ijkl}  &= (Q-1)\delta^{ab}\delta^{cd}P^S_{ijkl} \,,\\
{P^X}^{\lsp abcd}_{\lsp ijkl}  &= (Q-1) \left(\delta^{abcd}-\frac{1}{n}\delta^{ab}\delta^{cd} \right)P^S_{ijkl} \,,\\
{P^Z}^{\lsp abcd}_{\lsp ijkl}  &= \left(\delta^{ac}\delta^{bd} -\delta^{abcd}\right) P_{ik}P_{jl}+ \left(\delta^{ad}\delta^{bc} -\delta^{abcd}\right) P_{il}P_{jk} \,,\\
{P^B}^{\lsp abcd}_{\lsp ijkl}  &= \frac{1}{2}\left(\delta^{ac}\delta^{bd} -\delta^{abcd}\right) P_{ik}P_{jl}-\frac{1}{2} \left(\delta^{ad}\delta^{bc} -\delta^{abcd}\right)P_{il}P_{jk} \,,\\
{P^\phi}^{\lsp abcd}_{\lsp ijkl}  &=2\delta^{abcd}P^V_{ijkl} \,,\\
{P^A}^{\lsp abcd}_{\lsp ijkl}  &=2\delta^{abcd}P^A_{ijkl} \,,\\
{P^T}^{\lsp abcd}_{\lsp ijkl}  &= \delta^{abcd}P^T_{ijkl}\,.
\end{aligned}
\label{moresinglecorrprojectors}
\end{equation}
Notice that some of the above tensors are not unit normalised, this is for future convenience when considering mixed correlator systems. In order to obtain the sum rules from these tensor structures it is convenient to contract the correlator $\langle \phi^a_i \phi^b_j \phi^c_k \phi^d_l \rangle$ with vectors $v_i$ satisfying $\sum_i v_i=0$. This eliminates all factors of $\omega$ without modifying the sum rules. Thus one can obtain the sum rules simply by reading off the coefficients of the Kronecker deltas. In parts of this work, we use conventions that omit the $\omega$'s from the beginning for practical purposes.

\subsection{\texorpdfstring{$\phi$}{phi} Single Correlator Sum Rules}
\label{phisumrules}
We obtain the sum rules for the crossing equation $\langle \phi^a_i \phi^b_j \phi^c_k \phi^d_l \rangle = \langle  \phi^c_k \phi^b_j \phi^a_i \phi^d_l \rangle$ by reading the coefficients of Eq.~\eqref{crossingformula}, with respect to $\delta^{ac}\delta^{bd}\delta_{ik}\delta_{jl}$, $\delta^{ad}\delta^{bc}\delta_{il}\delta_{jk}$, $\delta^{ab}\delta^{cd}\delta_{ij}\delta_{kl}$, $\delta^{abcd}\delta_{ij}\delta_{kl}$, $\delta^{abcd}\delta_{ik}\delta_{jl}$, $\delta^{abcd}\delta_{il}\delta_{jk}$, and $\delta^{abcd}\delta_{ijkl}$. Notice that in the special case of $Q=3$ the four index Kronecker delta $\delta_{ijkl}$ can be written as a function of $\delta_{ijk}\omega_l$, $\delta_{ij}\delta_{kl}$, $\delta_{ij}\omega_i \omega_j$ and $\omega_i \omega_j \omega_k \omega_l$ and their index permutations. Concretely, $\delta_{ijkl}=\frac{1}{3}\omega_{(i} \delta_{jkl)}+\frac{1}{6}\delta_{(ij}\delta_{kl)}-\frac{1}{6}\omega_{(i}\omega_{j}\delta_{kl)}+\frac{1}{6}\omega_i \omega_j \omega_k \omega_l $. For $n \geq 2$ and $Q = 3$ the sum rules read
\begin{equation}
\begin{aligned}
0&=n(Q-1) F^-_{S}-2F^-_{A}-2(n-1)F^-_{B}\,,\\
0&= (Q-1)F^-_{X}-2F^-_{A}+2F^-_{B}\,,\\
0&= 2F^-_{V}+2F^-_{A}\,,\\
0&= 2F^-_{Z}+F^-_{B}\,,\\
0&= n(Q-1)F^+_{S}-4F^+_{V}+2F^+_{A}-2(n-1)F^+_{Z}+(n-1)F^+_{B}\,,\\
0&= (Q-1)F^+_{X}-4F^+_{V}+2F^+_{A}+2F^+_{Z}-F^+_{B}\,,
\end{aligned}
\end{equation}
whereas for $n \geq 2$ and $Q > 3$ they read 
\begin{equation}
\resizebox{\textwidth}{!}{\(
\begin{aligned}
0&= n(Q-1)F^-_{S}-2(Q-2)F^-_{A}-(Q-1)(n-1)F^-_{B}\,,\\
0&= (Q-1)F^-_{X}-2(Q-2)F^-_{A}+(Q-1)F^-_{B}\,,\\
0&= 2F^-_{V}+2F^-_{A}\,,\\
0&= F^-_{T}+2F^-_{A}\,,\\
0&= 2F^-_{Z}+F^-_{B}\,,\\
0&= n(Q-1)F^+_{S}-\frac{2(Q-1)}{Q-2}F^+_{V}-\frac{Q(Q-3)}{2(Q-2)}F^+_{T}+(Q-1)F^+_{A}-(Q-1)(n-1)F^+_{Z}+\frac{(Q-1)(n-1)}{2}F^+_{B}\,,\\
0&= (Q-1)F^+_{X}-\frac{2(Q-1)}{Q-2}F^+_{V}-\frac{Q(Q-3)}{2(Q-2)}F^+_{T}+(Q-1)F^+_{A}+(Q-1)F^+_{Z}-\frac{Q-1}{2}F^+_{B}\,.
\end{aligned}
\)}
\end{equation}
Notice that the former ones coincide with those for $S_n\ltimes (O(2))^n$, up to adding/subtracting rows and redefining OPE coefficients, originally derived in~\citep{Stergiou:2019dcv}. This is expected due to the group-subgroup relationship between $O(2)$ and $S_3$. Indeed, after removing factors of $\omega$,\footnote{Which as discussed above, does not affect the sum rules.} the tensor structures for both groups, with respect to this specific correlator, coincide.

\subsection{\texorpdfstring{$X$}{X} Single Correlator Sum Rules}
\label{phiprojectors}
To start off, we remind the reader that the $X\times X$ OPE decomposes as
\begin{equation}
X \times X \sim S+X+\overline{XX}
\end{equation}
for $n=3$ and as
\begin{equation}
X \times X \sim S+X+XX+\overline{XX}
\end{equation}
for $n\geq 4$. 
In order to eventually define sum rules mixing $\phi^a_i$ and $X^{ab}_{ij}$ as externals it is convenient to define some OPE conventions. Tensors in OPEs can be defined modulo numerical coefficients, which can be absorbed into the OPE coefficients. Different choices correspond to different conventions, which are of course all equivalent. In particular, in order to obtain the projectors in Eq.~\eqref{moresinglecorrprojectors} we have already implicitly defined
\begin{equation}
    \phi^a_i \times \phi^b_j \sim \lambda_{VVS} \delta^{ab}\left(\delta_{ij}-\frac{1}{Q}\omega_i \omega_j\right) S + \lambda_{VVX} X^{ab}_{ij}  + \cdots\,,
\end{equation}
where the dots refer to other irreps whose exact convention is not currently important. In our convention, $\langle X^{ab}_{ij} X^{cd}_{kl}\rangle \sim {P^X}^{\lsp abcd}_{\lsp ijkl}$, with $P^X$ as defined in Eq.~\eqref{moresinglecorrprojectors}. Consequently, associativity of the OPE forces
\begin{equation}
    \phi^a_i \times X^{bc}_{jk} \sim \lambda_{VXV} {P^X}^{\lsp adbc}_{\lsp iljk} \phi^d_l+\cdots
\end{equation}
Lastly, we define
\begin{equation}
    X^{ab}_{ij} \times X^{cd}_{kl} \sim \lambda_{XXS}{P^X}^{\lsp abcd}_{\lsp ijkl} S + \lambda_{XXX}\frac{1}{4} {P^X}^{\lsp abu_1 u_2}_{\lsp ij d_1 d_2} {P^X}^{\lsp cd u_1 u_3}_{\lsp kl d_1 d_3} {P^X}^{\lsp ef u_2 u_3}_{\lsp mn d_2 d_3} X^{ef}_{mn} +\cdots
\end{equation}
The above considerations fix the sum rules $\langle X^{ab}_{ij} X^{cd}_{kl} X^{ef}_{mn} X^{gh}_{op}\rangle $ $= \langle  X^{ef}_{mn} X^{cd}_{kl} X^{ab}_{ij} X^{gh}_{op} \rangle$ for $n=3$ to be
\begin{equation}
\begin{aligned}
0&=\frac{1}{6}F^-_{X}+\frac{1}{2}F^-_{\overline{XX}}\,,\\
0&= F^-_{S}-\frac{1}{2}F^-_{\overline{XX}}\,,\\
0&= F^+_{S}-\frac{2}{6}F^+_{X} +\frac{1}{2}F^+_{\overline{XX}}\,,
\end{aligned}
\end{equation}
which have been derived using
\begin{equation}
\begin{aligned}
 {P^S}^{\lsp abcdefgh}_{\lsp ijklmnop}&={P^X}^{\lsp abcd}_{\lsp ijkl} {P^X}^{\lsp efgh}_{\lsp mnop}\,,\\
{P^{X}}^{\lsp abcdefgh}_{\lsp ijklmnop}&= \frac{1}{6}\left({P^X}^{\lsp abef}_{\lsp ijmn}{P^X}^{\lsp cdgh}_{\lsp klop} + {P^X}^{\lsp abgh}_{\lsp ijop}{P^X}^{\lsp efcd}_{\lsp mnkl}-{P^X}^{\lsp abcd}_{\lsp ijkl}{P^X}^{\lsp efgh}_{\lsp mnop}\right)\,,\\
{P^{\overline{XX}}}^{\lsp abcdefgh}_{\lsp ijklmnop}&= \frac{1}{2} {P^X}^{\lsp abef}_{\lsp ijmn} {P^X}^{\lsp cdgh}_{\lsp klop} -\frac{1}{2} {P^X}^{\lsp abgh}_{\lsp ijop} {P^X}^{\lsp efcd}_{\lsp mnkl}.
\end{aligned}
\end{equation}
where we have also used the simplifying identity for $\delta_{abcdefgh}$ when $n=3$, which expresses it in terms of Kronecker deltas with less indices, see Eq. 2.22 in ~\citep{Kousvos:2018rhl}. On the other hand, for $n \geq 4$ we obtain
\begin{equation}
\begin{aligned}
0&= (n-1) F^-_{S}-(n-2)F^-_{\overline{XX}}\,,\\
0&= \frac{(n-1)(n-2)}{n}F^-_{X}+F^-_{\overline{XX}}\,,\\
0&= F^-_{XX}+F^-_{\overline{XX}}\,,\\
0&= (n-1) F^+_{S}-\frac{(n-1)^2}{n}F^+_{X}-\frac{n(n-3)}{2(n-2)}F^+_{XX} +\frac{n-1}{2}F^+_{\overline{XX}}\,,
\end{aligned}
\end{equation}
by using
\begin{equation}
\resizebox{\textwidth}{!}{\(
\begin{aligned}
{P^S}^{\lsp abcdefgh}_{\lsp ijklmnop} &={P^X}^{\lsp abcd}_{\lsp ijkl} {P^X}^{\lsp efgh}_{\lsp mnop} \,,\\
{P^X}^{\lsp abcdefgh}_{\lsp ijklmnop} &= d^{abcdrs} d^{efghrs} P_{ij}P_{kl}P_{mn}P_{op}\,,\\
{P^{XX}}^{\lsp abcdefgh}_{\lsp ijklmnop}&= \frac{1}{2}\left( {P^X}^{\lsp abef}_{\lsp ijmn}{P^X}^{\lsp cdgh}_{\lsp klop} + {P^X}^{\lsp abgh}_{\lsp ijop}{P^X}^{\lsp efcd}_{\lsp mnkl} \right)- \frac{n}{n-2} {P^X}^{\lsp abcdefgh}_{\lsp ijklmnop}-\frac{1}{n-1} {P^S}^{\lsp abcdefgh}_{\lsp ijklmnop} \,,\\
{P^{\overline{XX}}}^{\lsp abcdefgh}_{\lsp ijklmnop} &= \frac{1}{2}\left( {P^X}^{\lsp abef}_{\lsp ijmn}{P^X}^{\lsp cdgh}_{\lsp klop} - {P^X}^{\lsp abgh}_{\lsp ijop}{P^X}^{\lsp efcd}_{\lsp mnkl} \right) \,,\\
\end{aligned}
\)}
\label{xsinglecorrprojectors}
\end{equation}
where $d^{abcdef}=\delta^{abcdef}-\frac{1}{n}\left(\delta^{abcd}\delta^{ef} +\delta^{abef}\delta^{cd} +\delta^{cdef}\delta^{ab} \right)+\frac{2}{n^2}\delta^{ab} \delta^{cd} \delta^{ef}  $. Notice that the above projectors, mirror those of $S_Q$ written earlier, by swapping $Q$ for $n$. This is expected since the operator $X^{ab}_{ij}$ transforms under $S_n \ltimes (S_Q)^n$ in the same way that the defining representation of $S_n$ transforms under $S_n$. Indeed these sum rules coincide (up to normalization conventions) with the ones derived in~\citep{Rong:2017cow}. In our treatment, the tensor structures carry ``redundant indices'' ($X^{ab}_{ij}\propto \delta^{abc}P_{ij} X^c$) which prove however useful when considering mixed correlator systems.

\subsection{\texorpdfstring{$\phi - X$}{phi-X} Mixed Correlator Sum Rules}
\label{phixsumrules}
The sum rules resulting from the $\langle \phi^a_i \phi^b_j X^{cd}_{kl} X^{ef}_{mn} \rangle = \langle X^{cd}_{kl} \phi^b_j \phi^a_i X^{ef}_{mn}  \rangle$ and $\langle \phi^a_i X^{cd}_{kl} \phi^b_j X^{ef}_{mn} \rangle = \langle \phi^b_j X^{cd}_{kl} \phi^a_i X^{ef}_{mn} \rangle$ crossing equations can be obtained using the following tensors
\begin{equation}
\begin{aligned}
    {P^S}^{\lsp abcdef}_{ijklmn} &= P_{ij} \delta^{ab} {P^X}^{\lsp cdef}_{\lsp klmn} \,,\\
    {P^X}^{\lsp abcdef}_{ijklmn} &= P_{ij}P_{kl}P_{mn} d^{abcdef}\,,
\end{aligned}
\end{equation}
which appear in the $\langle \phi^a_i \phi^b_j X^{cd}_{kl} X^{ef}_{mn} \rangle$ correlator, and 
\begin{equation}
\resizebox{\textwidth}{!}{\(
\begin{aligned}
    {P^V}^{\lsp abcdef}_{ijklmn}    &=P_{ij} P_{kl} P_{mn}\left( \delta^{abcdef}-\frac{1}{n}\left(\delta^{abcd}\delta^{ef} +\delta^{abef}\delta^{cd} \right)+\frac{1}{n^2}\delta^{ab} \delta^{cd} \delta^{ef} \right) \,,\\
    {P^{VX}}^{\lsp abcdef}_{ijklmn} &= P_{ij} P_{kl} P_{mn}\left(-\delta^{abcdef} +\frac{1}{n}\left(\delta^{abcd}\delta^{ef} +\delta^{abef}\delta^{cd} \right) +\frac{n-1}{n} \delta^{ab}\delta^{cdef}+ \frac{1}{n}\delta^{ab}\delta^{cd}\delta^{ef} \right)\,,
\end{aligned}
\)}
\end{equation}
which appear in the $\langle \phi^a_i X^{cd}_{kl} \phi^b_j X^{ef}_{mn}\rangle$ correlator. These can be checked to be compatible with the OPE definitions of the previous section. We obtain
\begin{equation}
0=F^-_{V}
\end{equation}
and
\begin{equation}
0=F^-_{XV}
\end{equation}
resulting from $\langle \phi^a_i X^{cd} \phi^b_j X^{ef} \rangle = \langle \phi^b_j X^{cd} \phi^a_i X^{ef} \rangle$. As well as
\begin{equation}
0=F^\mp_S \pm (-1)^\ell \, \frac{1}{n} F^\mp_{V} \pm (-1)^\ell \, \frac{1}{n}F^\mp_{XV}
\end{equation}
and 
\begin{equation}
0=F^\mp_X \pm (-1)^\ell \,  F^\mp_{V} \pm (-1)^\ell \, F^\mp_{XV}
\end{equation}
resulting from $\langle \phi^a_i \phi^b_j X^{cd} X^{ef} \rangle = \langle X^{cd} \phi^b_j \phi^a_i X^{ef}  \rangle$.

\subsection{Tensor Structures for \texorpdfstring{$\braket{Z_3 Z_3 Z_3 Z_3}$}{<Z3 Z3 Z3 Z3>}}
\label{z3tensors}
In order to make contact with previous work in the literature, i.e.~\citep{Dotsenko:1998gyp}, we will work specifically in the case $n=3$, which is also the smallest value of $n$ for which there is a non-trivial fixed point. This will considerably simplify the derivation of the sum rules. A field in the $Z_3$ irrep can be represented with three upper (replica) and three lower indices, ${Z_3}^{\lsp abc}_{\lsp ijk}$. Since by the definition of this representation, the replica indices must be different, for $n=3$ we can \emph{wlog} take $a=1$, $b=2$ and $c=3$. We will thus consider the following correlator $\langle {Z_3}^{\lsp 123}_{\lsp ijk}{Z_3}^{\lsp 123}_{\lsp lmn}{Z_3}^{\lsp 123}_{\lsp opr}{Z_3}^{\lsp 123}_{\lsp stu} \rangle$. The $Z_3 \times Z_3$ OPE decomposes as follows
\begin{equation}
Z_3 \times Z_3 \sim AA+VAA+AZ+S+A+V+VA+Z+Z_3 +AAA \,.
\end{equation}
Remembering that the lower indices can only ``talk'' to each other if they are in the same copy we obtain the following tensor structures
\begin{equation}
\begin{aligned}
{P^{AA}}_{ijklmnoprstu}& = P^A_{ilos}P^A_{jmpt}P^S_{knru}+P^A_{ilos}P^S_{jmpt}P^A_{knru}+P^S_{ilos}P^A_{jmpt}P^A_{knru}\,,\\
{P^{AAV}}_{ijklmnoprstu} & = P^A_{ilos}P^A_{jmpt}P^\phi_{knru}+P^A_{ilos}P^\phi_{jmpt}P^A_{knru}+P^\phi_{ilos}P^A_{jmpt}P^A_{knru}\,,\\
{P^{AZ}}_{ijklmnoprstu} & = P^A_{ilos}P^\phi_{jmpt}P^\phi_{knru}+P^\phi_{ilos}P^\phi_{jmpt}P^A_{knru}+P^\phi_{ilos}P^A_{jmpt}P^\phi_{knru}\,,\\
{P^{S}}_{ijklmnoprstu} & = P^S_{ilos}P^S_{jmpt}P^S_{knru} \,,\\
{P^{A}}_{ijklmnoprstu} & = P^A_{ilos}P^S_{jmpt}P^S_{knru}+P^S_{ilos}P^A_{jmpt}P^S_{knru}+P^S_{ilos}P^S_{jmpt}P^A_{knru}\,,\\
{P^{V}}_{ijklmnoprstu} & = P^\phi_{ilos}P^S_{jmpt}P^S_{knru}+P^S_{ilos}P^\phi_{jmpt}P^S_{knru}+P^S_{ilos}P^S_{jmpt}P^\phi_{knru}\,,\\
{P^{VA}}_{ijklmnoprstu} & = P^S_{ilos}P^\phi_{jmpt}P^A_{knru}+P^S_{ilos}P^A_{jmpt}P^\phi_{knru}+P^A_{ilos}P^S_{jmpt}P^\phi_{knru}\,,\\
& \quad +P^A_{ilos}P^\phi_{jmpt}P^S_{knru}+P^\phi_{ilos}P^A_{jmpt}P^S_{knru}+P^\phi_{ilos}P^S_{jmpt}P^A_{knru}\,,\\
{P^{Z}}_{ijklmnoprstu} & = P^\phi_{ilos}P^S_{jmpt}P^S_{knru}+P^S_{ilos}P^\phi_{jmpt}P^S_{knru}+P^S_{ilos}P^S_{jmpt}P^\phi_{knru}\,,\\
{P^{Z_3}}_{ijklmnoprstu} & = P^\phi_{ilos}P^\phi_{jmpt}P^\phi_{knru}\,,\\
{P^{AAA}}_{ijklmnoprstu} & =P^A_{ilos}P^A_{jmpt}P^A_{knru} \,,
\end{aligned}
\label{z3z3projectors}
\end{equation}
where the $P^S$, $P^A$ and $P^\phi$ on the right hand side of Eq.~\eqref{z3z3projectors} are those defined earlier in Eq.~\eqref{singlecorrprojectors}, but multiplied by a factor of $2$.
\subsection{\texorpdfstring{$Z_3$}{Z3} Single Correlator Sum Rules}
\label{z3sumrules}
To obtain the sum rules from the $\langle {Z_3}^{\lsp 123}_{\lsp ijk}{Z_3}^{\lsp 123}_{\lsp lmn}{Z_3}^{\lsp 123}_{\lsp opr}{Z_3}^{\lsp 123}_{\lsp stu} \rangle = \langle {Z_3}^{\lsp 123}_{\lsp opr}{Z_3}^{\lsp 123}_{\lsp lmn}{Z_3}^{\lsp 123}_{\lsp ijk}{Z_3}^{\lsp 123}_{\lsp stu} \rangle$ crossing equation we first make the following convenient substitutions
\begin{equation}
\begin{aligned}
P^A_{ilos} &= P^0_{ilos}-\frac{P^+_{ilos}-P^-_{ilos}}{2}\,,\\
P^V_{ilos} &= P^0_{ilos}+\frac{P^+_{ilos}-P^-_{ilos}}{2}-\frac{P^+_{ilos}+P^-_{ilos}}{2}\,,\\
P^S_{ilos} &= \frac{P^+_{ilos}+P^-_{ilos}}{2}\,,
\end{aligned}
\end{equation}
and similarly for $P^A_{jmpt}$, $P^A_{knru}$ and the rest. We have defined $P^0$ and $P^+$ to be even under crossing, and $P^-$ to be odd. Reading the coefficients of each possible combination of $P^+$, $P^-$ and $P^0$ we obtain the following sum rules
\begin{equation}
\begin{aligned}
0&=2 F^-_{AAA} + 6 F^-_{AAV} + 6 F^-_{AZ} + 2 F^-_{Z_3}\,,\\
0&= F^-_{AA} - F^-_{AAA} - 2 F^-_{AAV} - F^-_{AZ} + 
 2 F^-_{VA} + F^-_{Z}\,,\\
0&=  \frac{1}{2} F^-_{A} - F^-_{AA} +\frac{1}{2} F^-_{AAA} + \frac{1}{2} F^-_{AAV} +\frac{1}{2} F^-_{V} - F^-_{VA}\,,\\
0&= \frac{1}{2}  F^-_{A} + F^-_{AA} + 
\frac{1}{2} F^-_{AAA} - \frac{3}{2} F^-_{AAV} +\frac{1}{2} F^-_{V} - F^-_{VA} - 2 F^-_{Z} + 2 F^-_{Z_3}\,,\\
0&= -\frac{3}{4} F^-_{A} + \frac{3}{4} F^-_{AA} -\frac{1}{4} F^-_{AAA} +\frac{1}{4} F^-_{S}\,,\\
0&= \frac{1}{4} F^-_{A} -\frac{1}{4} F^-_{AA} -\frac{1}{4} F^-_{AAA} + F^-_{AAV} - F^-_{AZ} +\frac{1}{4} F^-_{S} \,,
- F^-_{V} + F^-_{Z}\,,
\end{aligned}
\end{equation}
and 
\begin{equation}
\begin{aligned}
0&=  F^+_{AA} + F^+_{AAA} - 3 F^+_{AZ} + 2 F^+_{VA} + F^+_Z - 
 2 F^+_{Z_3}\,,\\
0&= \frac{1}{2}  F^+_{A} -\frac{1}{2} F^+_{AAA} +\frac{1}{2} F^+_{AAV} + F^+_{AZ} +\frac{1}{2} F^+_{V} - F^+_{VA} - F^+_{Z}\,,\\
0&=   -\frac{1}{4} F^+_{A} -\frac{1}{4} F^+_{AA} +\frac{1}{4} F^+_{AAA} -\frac{1}{2} F^+_{AAV} +\frac{1}{4} F^+_{S} -\frac{1}{2} F^+_{V} + F^+_{VA}\,,\\
0&=\frac{3}{4}    F^+_{A} +\frac{3}{4} F^+_{AA} +\frac{1}{4} F^+_{AAA} -\frac{3}{2}  F^+_{AAV} + 3 F^+_{AZ} +\frac{1}{4} F^+_{S} -\frac{3}{2}  F^+_{V} - 3 F^+_{VA} + 
 3 F^+_{Z} - 2 F^+_{Z_3}\,.
\end{aligned}
\end{equation}

\subsection{Tensor Structures for \texorpdfstring{$\braket{\phi Z_3 \phi Z_3}$}{<phi Z3 phi Z3>}}
\label{phiz3tensors}
To study the tensor structures of the $\langle \phi Z_3 \phi Z_3 \rangle$ correlators, as in the preceding section we will plug in explicit index values for the upper indices. Without loss of generality, the two indipendent cases to consider are $\langle \phi^1_i  {Z_3}^{\lsp 123}_{\lsp jkl} \phi^1_{m} {Z_3}^{\lsp 123}_{\lsp nop}\rangle$ and $\langle \phi^1_i  {Z_3}^{\lsp 123}_{\lsp jkl} \phi^2_{m} {Z_3}^{\lsp 123}_{\lsp nop}\rangle$. The $\phi \times Z_3$ OPE decomposes as
\begin{equation}
\phi \times Z_3 \sim Z+AZ+VB+Z_3.
\end{equation} 
In order to understand what tensors will appear, let us look more carefully at how the $VB$ and $Z_3$ representations come about\footnote{We temporarily omit other representations from the product $\phi \times Z_3$.}
\begin{equation}
\begin{aligned}
\phi^1_i \times {Z_3}^{\lsp 123}_{\lsp jkl} &\sim C_{ijs} V^1_s Z^{23}_{kl}\\
&\sim C_{ijs}(V^1_s Z^{23}_{kl} +V^2_k Z^{13}_{sl}+V^3_l Z^{12}_{sk})\\
&\quad +C_{ijs}(V^1_s Z^{23}_{kl} -V^2_k Z^{13}_{sl})+C_{ijs}(V^1_s Z^{23}_{kl} -V^3_l Z^{12}_{sk})\\
&\sim C_{ijs}({Z_3}^{\lsp 123}_{\lsp skl}+V^3_l B^{12}_{sk}+V^2_k B^{13}_{sl})\,.
\end{aligned}
\end{equation}
Plugging this equation into the four-point correlator one can find the tensor structures for each irrep. For the first of the two correlators, namely $\langle \phi^1_i  {Z_3}^{\lsp 123}_{\lsp jkl} \phi^1_{m} {Z_3}^{\lsp 123}_{\lsp nop}\rangle$, we have
\begin{equation}
\begin{aligned}
{P^Z}_{ijklmnop} &= \delta_{ij}\delta_{mn}\delta_{ko}\delta_{lp}\,,\\
{P^{AZ}}_{ijklmnop} & =\left(\delta_{im}\delta_{jn}-\delta_{in}\delta_{jm}\right)\delta_{ko}\delta_{lp} \,,\\
{P^{VB}}_{ijklmnop} & =2\left(\delta_{im}\delta_{jn}+\delta_{in}\delta_{jm}-\frac{2}{m}\delta_{ij}\delta_{mn}\right) \delta_{ko}\delta_{lp} \,,\\
{P^{Z3}}_{ijklmnop} & = \left(\delta_{im}\delta_{jn}+\delta_{in}\delta_{jm}-\frac{2}{m}\delta_{ij}\delta_{mn}\right) \delta_{ko}\delta_{lp}\,.
\end{aligned}
\end{equation}
Notice that while the $VB$ and $Z_3$ irreps above have the same tensor structure up to a numerical prefactor, this prefactor can and will be different in other correlators. Thus, the overall tensor structure will be different, as it should. Moving on to the second correlator of interest, $\langle \phi^1_i  {Z_3}^{\lsp 123}_{\lsp jkl} \phi^2_{m} {Z_3}^{\lsp 123}_{\lsp nop}\rangle$, we have
\begin{equation}
\begin{aligned}
{P^{Z_3}}_{ijklmnop} &= C_{ijr}C_{mos}\delta_{sk}\delta_{rn}\delta_{lp} = C_{ijn}C_{mok}\delta_{lp}\,,\\
{P^{VB}}_{ijklmnop} &= -C_{ijr}C_{mos}\delta_{sk}\delta_{rn}\delta_{lp}= -C_{ijn}C_{mok}\delta_{lp}\,,
\end{aligned}
\end{equation}
where now the tensors corresponding to the two irreps differ by a minus sign. The $AZ$ and $Z$ irreps do not appear in this correlator. As an aside, note that these two tensor structures will give rise to a sum rule of the form $F^-_{Z_3}-F^-_{VB} =0 $. This sum rule is characteristic of ``symmetry breaking'': it appears because the irrep $VZ$ of $S_3 \ltimes (O(2))^3$ becomes reducible when breaking the symmetry down to $S_3 \ltimes (S_3)^3$, due to the existence of the invariant tensor $C_{ijk}$.\footnote{Which in turn is invariant due to the breaking of the $Z_2$ contained in $O(2)$, when breaking $O(2)$ to $S_3$.}
\subsection{Tensor Structures for \texorpdfstring{$\braket{\phi \phi Z_3 Z_3}$}{<phi phi Z3 Z3>}}
\label{phiz3tensors2}
Using similar reasoning to the preceding subsection, we only need to consider the correlators $\langle \phi^1_m \phi^1_i {Z_3}^{\lsp 123}_{\lsp jkl} {Z_3}^{\lsp 123}_{\lsp nop} \rangle$ and $\langle \phi^2_m \phi^1_i {Z_3}^{\lsp 123}_{\lsp jkl} {Z_3}^{\lsp 123}_{\lsp nop} \rangle$. The irreps that contribute to these correlators are the ones that are exchanged in both the $\phi \times \phi$ and the ${Z_3} \times Z_3$ OPE. These are $S$, $A$, $V$ and $Z$. The tensor structures obtained for $\langle \phi^1_m \phi^1_i {Z_3}^{\lsp 123}_{\lsp jkl} {Z_3}^{\lsp 123}_{\lsp nop} \rangle$ are 
\begin{equation}
\begin{aligned}
{P^S}_{mijklnop} &= \delta_{im}\delta_{jn} \delta_{ko}\delta_{lp} \,,\\
{P^A}_{mijklnop} &= \left( \delta_{in}\delta_{mj}-\delta_{ij}\delta_{mn}\right)\delta_{ko}\delta_{lp}\,,\\
{P^V}_{mijklnop} &= \left( \delta_{in}\delta_{mj}+\delta_{ij}\delta_{mn}-\delta_{im}\delta_{jn}\right)\delta_{ko}\delta_{lp}\,,
\end{aligned}
\end{equation}
whereas for  $\langle \phi^2_m \phi^1_i {Z_3}^{\lsp 123}_{\lsp jkl} {Z_3}^{\lsp 123}_{\lsp nop} \rangle$ we have
\begin{equation}
{P^Z}_{mijklnop} =C_{ijn}C_{mok}\delta_{lp}\,,
\end{equation}
which is found using $C_{ijk} \sim \langle \phi_i \phi_j \phi_k \rangle$.

\subsection{\texorpdfstring{$\phi - Z_3$}{phi-Z3} Mixed Correlator Sum Rules}
\label{phiz3sumrules}
Having worked out the tensor structures in the two previous subsections it is straightforward to obtain the sum rules. We start from the  $\langle \phi^1_i  {Z_3}^{\lsp 123}_{\lsp jkl} \phi^1_{m} {Z_3}^{\lsp 123}_{\lsp nop}\rangle= \langle \phi^1_m  {Z_3}^{\lsp 123}_{\lsp jkl} \phi^1_{i} {Z_3}^{\lsp 123}_{\lsp nop}\rangle$ crossing equation from which we obtain 
\begin{equation}
\begin{aligned}
0&= 2 F^{-}_{AZ} + 4 F^{-}_{VB} + 2 F^{-}_{Z3}\,,\\
0&= -F^{-}_{AZ} + 2 F^{-}_{VB} + F^{-}_{Z} + F^{-}_{Z3} - 2 F^{-}_{VB} -  F^{-}_{Z_3}\,,
\end{aligned}
\end{equation}
and
\begin{equation}
0=F^+_{AZ} + F^+_Z - 4 F^+_{VB}  - 2F^+_{Z_3} .
\end{equation}
Consequently, from the crossing equation $\langle \phi^1_i  {Z_3}^{\lsp 123}_{\lsp jkl} \phi^2_{m} {Z_3}^{\lsp 123}_{\lsp nop}\rangle= \langle \phi^2_m  {Z_3}^{\lsp 123}_{\lsp jkl} \phi^1_{i} {Z_3}^{\lsp 123}_{\lsp nop}\rangle$ one obtains
\begin{equation}
0=F^-_{Z_3}-F^-_{VB}.
\end{equation}
These exhaust the number of sum rules one can obtain from crossing equations of the schematic type $\langle \phi Z_3 \phi Z_3\rangle = \langle \phi Z_3 \phi Z_3\rangle$. We thus move on to crossing equations of the schematic type $\langle \phi \phi Z_3 Z_3\rangle = \langle Z_3 \phi \phi Z_3 \rangle$. These are $\langle \phi^1_m \phi^1_i {Z_3}^{\lsp 123}_{\lsp jkl} {Z_3}^{\lsp 123}_{nop}\rangle = \langle {Z_3}^{\lsp 123}_{\lsp jkl}  \phi^1_i  \phi^1_m {Z_3}^{\lsp 123}_{nop} \rangle$ and $\langle \phi^2_m \phi^1_i {Z_3}^{\lsp 123}_{\lsp jkl} {Z_3}^{\lsp 123}_{nop}\rangle = \langle {Z_3}^{\lsp 123}_{\lsp jkl}  \phi^1_i  \phi^2_m {Z_3}^{\lsp 123}_{nop} \rangle$. Notice that while we will omit the OPE coefficients with which exchanged operators appear, we have taken into account the minus signs that occur due to $\lambda_{\phi Z_3 O} =(-1)^\ell \, \lambda_{Z_3 \phi O}$, where $\ell$ is the spin of the exchanged operator $O$. From the first of the two aforementioned crossing equations we obtain
\begin{equation}
\begin{aligned}
0&=  (-1)^{1+\ell} \,  F^{-}_{AZ} + F^{-}_S - F^{-}_{A} + (-1)^\ell \, F^{-}_{Z}\,,\\ 
0&= (-1)^{1+\ell} \,  F^{-}_{AZ} + F^{-}_S + F^{-}_{A} - 2 F^{-}_V + 
   4 (-1)^\ell \, F^{-}_{VB} + 2 (-1)^\ell \, F^{-}_{Z_3} + (-1)^{1+\ell} \, F^{-}_{Z}\,,\\ 
   0&=(-1)^{\ell} \, F^{-}_{AZ} + F^{-}_{A} + F^{-}_V + 2 (-1)^\ell \, F^{-}_{VB} + (-1)^\ell \, F^{-}_{Z_3}\,,
\end{aligned}
\end{equation}
and
\begin{equation}
\begin{aligned}
0&=(-1)^\ell \, F^{+}_{AZ} + F^{+}_S - F^{+}_{A} + (-1)^{1+\ell} \,  F^{+}_{Z}\,,\\
0&= (-1)^\ell \, F^{+}_{AZ} + 
   F^{+}_S + F^{+}_{A} - 2 F^{+}_V - 4 (-1)^\ell \, F^{+}_{VB} - 
   2 (-1)^\ell \, F^{+}_{Z3} + (-1)^\ell \, F^{+}_{Z}\,,\\
0&=    (-1)^{1+\ell} \,  F^{+}_{AZ} + F^{+}_{A} + F^{+}_{V} - 
 2 (-1)^\ell \, F^{+}_{VB} + (-1)^{1+\ell} \,  F^{+}_{Z_3}\,,
\end{aligned}
\end{equation}
where operators in the $Z$, $AZ$, $Z_3$ and $VB$ representations appear with $\lambda_{\phi Z_3 O}^2$ and $S$, $A$ and $V$ and $V_B$ appear with $\lambda_{\phi \phi O }\lambda_{Z_3 Z_3 O}$. From the second of the two crossing equations we obtain
\begin{equation}
0=(-1)^{1+\ell} \,  F^{-}_{VB} + F^{-}_Z + (-1)^\ell \, F^{-}_{Z_3}
\end{equation}
and
\begin{equation}
0=(-1)^{\ell} \, F^{+}_{VB} + F^{+}_Z + (-1)^{1+\ell} \,  F^{+}_{Z_3}\,,
\end{equation}
where operators in the $Z_3$ and $VB$ representations appear with $\lambda_{\phi Z_3 O}^2$ and $Z$ now appears with $\lambda_{\phi \phi O }\lambda_{Z_3 Z_3 O}$. We emphasise that $Z$ appears with a different combination of OPE coefficients in this crossing equation compared to the previous one.

\section{Numerical Implementation}
\label{numericalparameters}
All the results we presented in this text have been computed with the use of {\bf simpleboot}.\footnote{\href{https://gitlab.com/bootstrapcollaboration/simpleboot}{https://gitlab.com/bootstrapcollaboration/simpleboot}. See \href{https://perimeterinstitute.ca/events/mini-course-numerical-conformal-bootstrap}{here} for lectures on this software.}  The parameters we used in this software are $d=2$, $\Lambda=19$, $\kappa=14$, $r_N=56$ and  $\ell\ped{set} = \{0,\ldots,26, 49, 50\}$, the last of which corresponds to the spins of exchanged operators. In the case of Figure~\ref{fig:Xspin0PhiX11vs19}, and only there, we also present an allowed region obtained with the parameters  $d=2$, $\Lambda=11$, $\kappa=12$, $r_N=48$ and  $\ell\ped{set} = \{0,\ldots,20, 49, 52\}$. While calling {\bf SDPB}~\citep{Simmons-Duffin:2015qma,Landry:2019qug}, we used the parameters in Table~\ref{tab:sdpb-params}. During the early stages of this work, we also performed tests using {\bf PyCFTBoot}~\citep{Behan:2016dtz} and {\bf qboot}~\citep{Go:2020ahx}. 

\begin{table}[H]
    \centering
    \begin{tabular}{cc}
        \toprule
        parameter & value \\ \midrule
        \texttt{precision} & $1024$ \\
        \texttt{maxIterations} & $1000$ \\
        \texttt{maxComplementarity} & $10^{70}$ \\
        \texttt{dualityGapThreshold} & $10^{-40}$ \\
        \texttt{primalErrorThreshold} & $10^{-60}$ \\
        \texttt{dualErrorThreshold} & $10^{-60}$ \\
        \texttt{initialMatrixScalePrimal} & $10^{20}$ \\
        \texttt{initialMatrixScaleDual} & $10^{20}$ \\
        \texttt{detectPrimalFeasibleJump} & \texttt{True} \\
        \texttt{detectDualFeasibleJump} & \texttt{True} \\
        \bottomrule
    \end{tabular}
    \caption{SDPB parameters.}
    \label{tab:sdpb-params}
\end{table}

\section{Quasi Primaries of the Critical 3-state Potts Model}
\label{AppQuasiPrimaries}
The quasi-primary spectrum of the $Q = 3$ (bi-)critical Potts model can be obtained from the partition function by standard character techniques. The CFT has a finite number of Virasoro primaries reported in Table~\ref{tab:potts3-virasoro-primaries} along with the $S_{3}$ representation where they belong to. The modular invariant partition function is a particular bilinear combination of ``left'' and ``right''  characters of singular Verma modules~\citep{DiFrancesco:1997nk}:
\begin{equation}
    \mathcal{Z}(q, \bar{q}) = \sum_{r=1,2}\left(|\chi_{r,1}(q) + \chi_{r,5}(q)|^{2} + 2 |\chi_{r,3}(q)|^{2}\right)\,,
\end{equation}
where
\begin{equation}
    \begin{aligned}
        \chi_{r,s}(q) &= K_{\lambda_{r,s}}(q) - K_{\lambda_{r,s}}(q) \,,\\
        K_{\lambda}(q) &= \frac{1}{\eta(q)} \sum_{n \in \mathbb{Z}} q^{\frac{(Nn +\lambda)^{2}}{2N}} \,,\\
    \end{aligned}
\end{equation}
and $\lambda_{r,s} = p r - p' s$, $N = 2 p p'$, $\eta(q)$ is the Dedekin $\eta$-function. For the 3-state Potts model $p = 6, p' = 5$.

\begin{table}[t]
    \centering
    \begin{tabular}{ccccc}
        \toprule
        operator & $\Delta$ & $\ell$ & $((r,s);(\bar{r},\bar{s}))$ & $S_{3}$ irrep \\ \midrule
        $\id$ & 0 & 0 & $((1,1);(1,1))$ & singlet \\
        $\epsilon$ & $4/5$ & 0 & $((2,1);(2,1))$ & singlet \\
        $\sigma$ & $2/15$ & 0 & $((2,3);(2,3))$ & fundamental \\
        $\sigma'$ & $4/3$ & 0 & $((1,3);(1,3))$ & fundamental \\
        $\gamma$ & $9/5$ & 1 & $((2,1);(2,5))\oplus((2,5);(2,1))$ & antisymmetric \\
        $W$ & $3$ & 3 & $((1,1);(1,5))\oplus((1,5);(1,1))$ & antisymmetric \\
        $\epsilon'$ & $14/5$ & 0 & $((2,5);(2,5))$ & singlet \\
        $\epsilon''$ & $6$ & 0 & $((1,5);(1,5))$ & singlet \\
        \bottomrule
    \end{tabular}
    \caption{Virasoro primaries of the modular invariant critical 3-state Potts model.}
    \label{tab:potts3-virasoro-primaries}
\end{table}

The above partition function can be split in sectors of $S_{3}$ representation as follows
\begin{equation}
    \begin{aligned}
        \mathcal{Z}\ped{sing}(q,\bar{q}) &= |\chi_{1,1}(q)|^{2} + |\chi_{2,1}(q)|^{2} + |\chi_{2,5}(q)|^{2} + |\chi_{1,2}(q)|^{2} \,,\\  
        \mathcal{Z}\ped{fund}(q,\bar{q}) &= |\chi_{2,3}(q)|^{2} + |\chi_{1,3}(q)|^{2}\,,\\
        \mathcal{Z}\ped{asym}(q,\bar{q}) &= \chi_{2,1}(q)\chi_{2,5}(\bar{q})+\chi_{2,5}(q)\chi_{2,1}(\bar{q}) + \chi_{1,1}(q)\chi_{1,5}(\bar{q})+\chi_{1,5}(q)\chi_{1,1}(\bar{q})\,. \\
    \end{aligned}
\end{equation}
Each sector can be then decomposed in characters of the global conformal group to read off the quasi-primary spectrum
\begin{equation}
    \mathcal{Z}_{R}(s, x) = \mathcal{Z}_{R}(q = s x, \bar{q} = s/x) = s^{-\frac{c}{12}}\sum_{(\Delta, \ell)} p_{R, \Delta, \ell} \, \chi_{\Delta, \ell}(s, x) \,,
\end{equation}
where $R = \text{sing},\text{fund},\text{asym}$, $c = 4/5$ is the central charge, $p$ denotes the degeneracy of the quasi-primary and the characters are given by
\begin{equation}
    \begin{aligned}
        \chi_{\Delta, 0}(s, x) &= \frac{s^{\Delta}}{(1-s x)(1-s/x)} & (\Delta > 0)\,,\\
        \chi_{\Delta, \ell}(s, x) &= \frac{s^{\Delta}}{(1-s x)(1-s/x)} \left(x^{\ell} + \frac{1}{x^{\ell}}\right) & (\Delta > \ell, \ell \geq 1) \, ,\\
        \chi_{0, 0}(s, x) &= 1\,,\\
        \chi_{\ell, \ell}(s, x) &= \frac{1}{(1-s x)(1-s/x)} \left[s^{\ell}\left(x^{\ell} + \frac{1}{x^{\ell}}\right) - s^{\ell+1}\left(x^{\ell-1} + \frac{1}{x^{\ell-1}}\right)\right]\,.
    \end{aligned}    
\end{equation}
In Table~\ref{tab:potts3-quasi-primaries} we report the scaling dimensions and spins of the first 19 operators for each $S_3$ sector. 

These are of use in our work since both at the decoupled limit, as well as the the large-$n$ limit, the dimensions of operators in the replica theories are completely determined from those of a single copy. Hence, they provide a minimal intuition for what gap assumptions one may impose on the theory.

We observe that the operators $O_{S,1}$, $O_{V,1}$ and $O_{A,0}$ have particularly large gaps. While performing our numerical bootstrap bounds we experimented with imposing large gaps in these sectors, but this unfortunately did not produce an island, or alternatively, a peninsula which appears to be converging towards an island. In particular, we experimented with using $O_{V,0}$ and $O_{V^\prime,0}$ as externals in a single Potts model,\footnote{Which totals to $21$ sum rules that can be obtained via {\bf autoboot}~\citep{Go:2019lke}.} since this exchanges all three operators in question, and imposed $O_{S,1} \geq 7.5 $, $O_{V,1} \geq 3.5$, $O_{A,0} \geq 6.5$ and $O_{V^{\prime \prime},0 }\geq 4.1$ (while the exact value is $\Delta_{V^{\prime \prime},0}\simeq4.13$) at $\Lambda = 11$. 

For that reason, as also mentioned in the main text, throughout our work we aimed to instead impose rather conservative gap assumptions so that our bounds present somewhat ``universal'' (i.e.\ not too heavily gap-dependent) constraints on our theories of interest. This is especially important since we do not know anything about the anomalous dimensions that the aforementioned operators obtain at the coupled replica fixed point.

\begin{table}[t]
    \renewcommand{\arraystretch}{0.68}
    \centering
    \subcaptionbox{singlet}[0.2\textwidth]{
        \begin{NiceTabular}{ccc}
            \CodeBefore
            \rowcolor{cyan!25}{2,3,4,6}
            \Body
            \toprule
            $\ell$ & $\Delta$ & $p$ \\\midrule
            0 & 0 & 1 \\ 
            & $0.8$ & 1 \\ 
            & $2.8$ & 1 \\ 
            & 4 & 1 \\ 
            & 6 & 1 \\
            & $6.8$ & 2 \\ 
            1 & $7.8$ & 1 \\ 
            2 & 2 & 1 \\ 
            & $4.8$ & 1 \\ 
            & 6 & 1 \\ 
            3 & $3.8$ & 1 \\ 
            4 & 4 & 1 \\ 
            & $4.8$ & 1 \\ 
            & $6.8$ & 2 \\ 
            5 & $5.8$ & 1 \\ 
            & $7.8$ & 1 \\ 
            6 & 6 & 2 \\ 
            & $6.8$ & 2 \\
            7 & $7.8$ & 2 \\ \bottomrule
        \end{NiceTabular}
    }
    \subcaptionbox{fundamental}[0.2\textwidth]{
        \begin{NiceTabular}{ccc}
            \CodeBefore
            \rowcolor{cyan!25}{2,3}
            \Body
            \toprule
            $\ell$ & $\Delta$ & $p$ \\ \midrule
            0 & $0.13$ & 1 \\ 
            & $1.33$ & 1 \\ 
            & $4.13$ & 1 \\ 
            & $5.33$ & 1 \\ 
            & $6.13$ & 1 \\ 
            1 & $5.13$ & 1 \\ 
            & $7.13$ & 2 \\ 
            2 & $2.13$ & 1 \\ 
            & $3.33$ & 1 \\ 
            & $6.13$ & 2 \\ 
            3 & $3.13$ & 1 \\ 
            & $7.13$ & 2 \\ 
            4 & $4.13$ & 2 \\ 
            & $5.33$ & 2 \\ 
            5 & $5.13$ & 2 \\ 
            & $6.33$ & 1 \\ 
            6 & $6.13$ & 3 \\ 
            & $7.33$ & 3 \\ 
            & $7.13$ & 4 \\
            \bottomrule
        \end{NiceTabular}
    }
    \subcaptionbox{antisymmetric}[0.2\textwidth]{
        \begin{NiceTabular}{ccc}
            \CodeBefore
            \rowcolor{cyan!25}{3,8}
            \Body
            \toprule
            $\ell$ & $\Delta$ & $p$ \\ \midrule
            0 & $6.8$ & 2 \\ 
            1 & $1.8$ & 1 \\ 
            & $5$ & 1 \\ 
            & $7$ & 1 \\ 
            & $7.8$ & 1 \\ 
            2 & $4.8$ & 1 \\ 
            3 & $3$ & 1 \\ 
            & $3.8$ & 1 \\ 
            & $5.8$ & 1 \\ 
            & $7$ & 1 \\ 
            4 & $6.8$ & 1 \\ 
            5 & $5$ & 1 \\ 
            & $5.8$ & 2 \\ 
            & $7.8$ & 2 \\ 
            6 & $6$ & 1 \\ 
            & $6.8$ & 1 \\ 
            7 & $7$ & 1 \\ 
            & $7.8$ & 3 \\ 
            8 & $8$ & 1 \\
            \bottomrule
        \end{NiceTabular}
    }
    \caption{quasi-primary spectrum of critical 3-state Potts model. Coloured rows are the Virasoro primaries. The dimensions are all integers or rational ($x.13$ and $y.33$ should be read as $x + 2/15$ and $y + 1/3$).}
    \label{tab:potts3-quasi-primaries}
\end{table}

\section{Character Table}
\label{sec:CharTable}

\begin{table}[H]
\caption{Character table of $S_3 \ltimes (S_3)^3$, namely the value of the character $\chi_{\rho}([g])$ for each representation $\rho$ and for all the conjugacy classes $[g]$. The first row gives the sizes of the conjugacy classes.}
\label{tab:character-table} 
\aboverulesep=0ex
\belowrulesep=0ex
\resizebox{\textwidth}{!}{\begin{tabular}{ l | c c c c c c c c c c c c c c c c c c c c c c}
\toprule
c.c.s. & 1& 6& 12& 8& 27& 54& 72& 144& 18& 36& 36& 72& 162& 27& 9& 36& 36& 216& 54& 108& 54& 108 \\ \midrule
$S$  & 1 & 1 & 1 & 1 & 1 & 1 & 1 & 1 & 1 & 1 & 1 & 1 & 1 & 1 & 1 & 1 & 1 & 1 & 1 & 1 & 1 & 1 \\ 

$\overline{AAA}$ & 1 & 1 & 1 & 1 & 1 & 1 & 1 & 1 & $-1$ & $-1$ & $-1$ & $-1$ & $-1$ & $-1$ & $-1$ & $-1$ & $-1$ & $-1$ & 1 & 1 & 1 & 1 \\ 

$\overline{XX}$ & 1 & 1 & 1 & 1 & 1 & 1 & 1 & 1 & $-1$ & $-1$ & $-1$ & $-1$ & $-1$ & 1 & 1 & 1 & 1 & 1 & $-1$ & $-1$ & $-1$ & $-1$ \\ 

$AAA$&1 & 1 & 1 & 1 & 1 & 1 & 1 & 1 & 1 & 1 & 1 & 1 & 1 & $-1$ & $-1$ & $-1$ & $-1$ & $-1$ & $-1$ & $-1$ & $-1$ & $-1$ \\
$\overline{AA}A$&2 & 2 & 2 & 2 & 2 & 2 & $-1$ & $-1$ & 0 & 0 & 0 & 0 & 0 & $-2$ & $-2$ & $-2$ & $-2$ & 1 & 0 & 0 & 0 & 0 \\

$X$& 2 & 2 & 2 & 2 & 2 & 2 & $-1$ & $-1$ & 0 & 0 & 0 & 0 & 0 & 
2 & 2 & 2 & 2 & $-1$ & 0 & 0 & 0 & 0 \\ 

$XA$&3 & 3 & 3 & 3 & $-1$ & $-1$ & 0 & 0 & $-1$ & $-1$ & $-1$ & $-1$ & 1 & $-3$ & 1 & 1 & 1 & 0 & 1 & 1 & $-1$ & $-1$ \\ 

$\overline{AA}$&3 & 3 & 3 & 3 & $-1$ & $-1$ & 0 & 0 & $-1$ & $-1$ & $-1$ & $-1$ & 1 
& 3 & $-1$ & $-1$ & $-1$ & 0 & $-1$ & $-1$ & 1 & 1 \\ 

$A$&3 & 3 & 3 & 3 & $-1$ & $-1$ & 0 & 0 & 1 & 1 & 1 & 1 & $-1$ & $-3$ & 1 & 1 & 1 & 0 & $-1$ & $-1$ & 1 & 1 \\

$AA$&3 & 3 & 3 & 3 & $-1$ & $-1$ & 0 & 0 & 1 & 1 & 1 & 1 & $-1$ & 3 & $-1$ & $-1$ & $-1$ & 0 & 1 & 1 & $-1$ & $-1$ \\ 

$V\overline{AA}$&6 & 3 & 0 & $-3$ & 2 & $-1$ & 0 & 0 & $-2$ & $-2$ & 1 & 1 & 0 & 0 & $-4$ & $-1$ & 2 & 0 & 0 & 0 & 2 & $-1$ \\ 

$XV$&6 & 3 & 0 & $-3$ & 2 & $-1$ & 0 & 0 & $-2$ & $-2$ & 1 & 1 & 0 & 0 & 4 & 1 & $-2$ & 0 & 0 & 0 & $-2$ & 1 \\ 

$VAA$&6 & 3 & 0 & $-3$ & 2 & $-1$ & 0 & 0 & 2 & 2 & $-1$ & $-1$ & 0 & 0 & $-4$ & $-1$ & 2 & 0 & 0 & 0 & $-2$ & 1 \\ 

$V$&6 & 3 & 0 & $-3$ & 2 & $-1$ & 0 & 0 & 2 & 2 & $-1$ & $-1$ & 0 & 0 & 4 & 1 & $-2$ & 0 & 0 & 0 & 2 & $-1$ \\ 

$B_3$&8 & $-4$ & 2 & $-1$ & 0 & 0 & 2 & $-1$ & $-4$ & 2 & 2 & $-1$ & 0 & 0 & 0 & 0 & 0 & 0 & 0 & 0 & 0 & 0 \\ 

$Z_3$&8 & $-4$ & 2 & $-1$ & 0 & 0 & 2 & $-1$ & 4 & $-2$ & $-2$ & 1 & 0 & 0 & 0 & 0 & 0 & 0 & 0 & 0 & 0 & 0 \\ 

$VA$&12 & 6 & 0 & $- 6$ & $-4$ & 2 & 0 & 0 & 0 & 0 & 0 & 0 & 0 & 0 & 0 & 0 & 0 & 0 & 0 & 0 & 0 & 0 \\ 

$AB$&12 & 0 & $-3$ & 3 & 0 & 0 & 0 & 0 & $-2$ & 1 & $-2$ & 1 & 0 & 0 & $-4$ & 2 & $-1$ & 0 & 2 & $-1$ & 0 & 0 \\ 

$B$&12 & 0 & $-3$ & 3 & 0 & 0 & 0 & 0 & $-2$ & 1 & $-2$ & 1 & 0 & 0 & 4 & $-2$ & 1 & 0 & $-2$ & 1 & 0 & 0 \\

$AZ$&12 & 0 & $-3$ & 3 & 0 & 0 & 0 & 0 & 2 & $-1$ & 2 & $-1$ & 0 & 0 & $-4$ & 2 & $-1$ & 0 & $-2$ & 1 & 0 & 0 \\ 

$Z$&12 & 0 & $-3$ & 3 & 0 & 0 & 0 & 0 & 2 & $-1$ & 2 & $-1$ & 0 & 0 & 
4 & $-2$ & 1 & 0 & 2 & $-1$ & 0 & 0 \\ 

$VB$&16 & $-8$ & 4 & $-2$ & 0 & 0 & $-2$ & 1 & 0 & 0 & 0 & 0 & 0 & 0 & 0 & 0 & 0 & 0 & 0 & 0 & 0 & 0  \\
\bottomrule
\end{tabular}}
\end{table}

\section{Sum Rules from Finite Group Characters}
\label{discreteformulas}

We describe an alternative method to derive the sum rules for specific correlators. This allows us to obtain the sum rules directly from the character table of a given discrete group (accessible e.g.\ with \href{https://www.gap-system.org/}{{\bf GAP}}~\citep{GAP4}). This method has provided an additional check for part of the sum rules derived using the projectors/``tensor'' methods. 

Let \( G \) be a finite group, \( V \) a representation and \( R \) an irreducible representation. The projection map \( V \to R \) is known to be given as an average of the representation \( V \) over the group weight by the character of the representation \( R \)
\begin{equation}
    (P^{V}_{R})_{ij} = \frac{d_{R}}{|G|} \sum_{g} \chi_{R}(g)^{*} \, V(g)_{ij}\,,
\end{equation}
where $d_{R}$ signifies the dimension of the representation. We henceforth assume that all matrix representations are put in a unitary form, but the result will be basis-independent. 

Thus, the complex conjugate of the projection matrix is
\begin{equation}
    (P^{V}_{R})_{ij}^{*} = (P^{V}_{R})_{ji}\,.
\end{equation}
Hence, the projectors satisfy the following orthogonality relation (sum over repeated indices)
\begin{equation}
    (P^{V}_{R})_{ji}^{*} (P^{V}_{S})_{jk} = (P^{V}_{R})_{ij} (P^{V}_{S})_{jk} = \delta_{RS} (P^{V}_{R})_{jk}\,,
\end{equation}
and are normalised to
\begin{equation}
    \tr[P^{V}_{R}] = \frac{d_{R}}{|G|} \sum_{g} \chi_{R}(g)^{*} \chi_{V}(g) = d_{R} \braket{\chi_{R}, \chi_{V}} = d_{R} n_{R,V}\,,
\end{equation}
where \( n_{R,V} \) is multiplicity of the \( R \) representation in the decomposition of \( V \) into irreps and \( \braket{\chi, \chi'} \) is the usual scalar product of characters.

Moreover, if \( V \) is irreducible, then the projector is the identity if \( R = V \) and zero otherwise by the usual orthogonality relation of irreducible representations
\begin{equation}
    (P^{V}_{R})_{ij} = \frac{d_{R}}{|G|} \sum_{g}\chi_{R}(g)^{*} \, V(g)_{ij} = \frac{d_{R}}{|G|} \sum_{g}R(g)_{kk}^{*} V(g)_{ij} = \delta_{RV} \delta_{ij}.
\end{equation}
If \( V = V_{1} \otimes V_{2} \), then the projection to \( R \in V_{1} \otimes V_{2} \) is
\begin{equation}
    (P^{V_{1}\otimes V_{2}}_{R})_{ijkl} = \frac{d_{R}}{|G|} \sum_{g} \chi_{R}(g)^{*} \, V_{1}(g)_{ij} V_{2}(g)_{kl}.
\end{equation}
This corresponds to the contraction of two Clebsh-Gordans (CG) for the decomposition \( V_{1} \otimes V_{2} \to R \). Indeed if CG are defined by
\begin{equation}
    (V_{1}\otimes V_{2})(g)_{ij,kl} = V_{1}(g)_{ij} V_{2}(g)_{kl} = \sum_{R \in V_{1} \otimes V_{2}}\sum_{\ell=1}^{n_{R}}\cg{V_{1}}{i}{V_{2}}{k}{R}{n}_{\ell} \cg{V_{1}}{j}{V_{2}}{l}{R}{m}^{*}_{\ell} \, R(g)_{nm}
\end{equation}
where \( n_{R} \) is the degeneracy of the irrep \( R \) in the \( V_{1}\otimes V_{2} \) decomposition. Then
\begin{align*}
  (P^{V_{1} \otimes V_{2}}_{R})_{ijkl} &= \frac{d_{R}}{|G|} \sum_{g} \chi_{R}(g)^{*}\sum_{S \in V_{1} \otimes V_{2}} \sum_{\ell=1}^{n_{R}}\cg{V_{1}}{i}{V_{2}}{k}{S}{n}_{\ell} \cg{V_{1}}{j}{V_{2}}{l}{S}{m}^{*}_{\ell} \, S(g)_{nm} \\
                   &= \sum_{S \in V_{1} \otimes V_{2}} \sum_{\ell=1}^{n_{R}}\cg{V_{1}}{i}{V_{2}}{k}{S}{n}_{\ell} \cg{V_{1}}{j}{V_{2}}{l}{S}{m}^{*}_{\ell} \left(\frac{d_{R}}{|G|} \sum_{g} R(g)_{pp}^{*} S(g)_{nm}\right) \\
  &= \sum_{S \in V_{1} \otimes V_{2}}\sum_{\ell=1}^{n_{R}}\cg{V_{1}}{i}{V_{2}}{k}{S}{n}_{\ell} \cg{V_{1}}{j}{V_{2}}{l}{S}{m}^{*}_{\ell} \delta_{RS} \delta_{np} \delta_{mp}\,.
\end{align*}
Thus
\begin{equation}
    (P^{V_{1} \otimes V_{2}}_{R})_{ijkl} = \sum_{\ell=1}^{n_{R}}\cg{V_{1}}{i}{V_{2}}{k}{R}{n}_{\ell} \cg{V_{1}}{j}{V_{2}}{l}{R}{m}^{*}_{\ell}\,.
\end{equation}

Following the method described in~\citep{Go:2019lke}, one can write down the sum rules for correlators of $SO(d)$ scalars in the presence of some global symmetry in terms of contractions of CG. We shall then make use of the above formula to write the sum rules in terms of group characters. 

The method applies to 4-point functions of type
\begin{equation}
    \braket{v_{i}(x_{1}) w_{j}(x_{2}) v_{k}(x_{3})^{*} w_{l}(x_{4})^{*}}\,,
\end{equation}
where \(v, w \) are scalar primaries in the \(V\) and \(W\) irreducible representations of the global symmetry group \(G\). They admit the following OPE decomposition in the \( (12)(34) \) channel
\begin{equation}
    \braket{v_{i}(x_{1}) w_{j}(x_{2}) v_{k}(x_{3})^{*} w_{l}(x_{4})^{*}} = \mathbf{K}_{12,34} \sum_{R \in V \otimes W} \cg{V}{i}{W}{j}{R}{n}\cg{V^{*}}{k}{W^{*}}{l}{R^{*}}{n} g_{R}^{vw, vw}(z, \bar{z})\,,
\end{equation}
where
\begin{equation}
    \mathbf{K}_{ij,kl} = \frac{1}{|x_{12}|^{\Delta_{i} + \Delta_{j}} |x_{34}|^{\Delta_{k} + \Delta_{l}}} \left(\frac{|x_{24}|}{|x_{14}|}\right)^{\Delta_{i}-\Delta_{j}} \left(\frac{|x_{14}|}{|x_{13}|}\right)^{\Delta_{k}-\Delta_{l}}
\end{equation}
and $u = z \bar{z} = x_{12}^{2} x_{34}^{2}/(x_{13}^{2} x_{24}^{2}), v = (1-z) (1-\bar{z}) = x_{23}^{2} x_{14}^{2}/(x_{13}^{2} x_{24}^{2})$ are the usual cross rations ($x_{ij} = x_{i} - x_{j}$). We have further assumed that the irrep decomposition of \( V \otimes W \) is degeneracy-free. 

Observing that
\begin{equation}\label{eq:cg-complex-conjugate}
    \cg{V^{*}}{k}{W^{*}}{l}{R^{*}}{n} = \cg{V}{k}{W}{l}{R}{n}^{*}\,,
\end{equation}
we recognise the projector \( V\otimes W \to R \) 
\begin{equation}
    \braket{v_{i}(x_{1}) w_{j}(x_{2}) v_{k}(x_{3})^{*} w_{l}(x_{4})^{*}} = \mathbf{K}_{12,34}\sum_{R \in V \otimes W} (P^{V \otimes W}_{R})_{ikjl} g_{R}^{vw, vw}(z, \bar{z})\,.
\end{equation}
If we perform the OPE in the \( (14)(32) \) channel instead we have
\begin{equation}
\resizebox{\textwidth}{!}{\(
    \braket{v_{i}(x_{1}) w_{j}(x_{2}) v_{k}(x_{3})^{*} w_{l}(x_{4})^{*}} = \mathbf{K}_{14,32}\sum_{S \in V \otimes W^{*}} \cg{V}{i}{W^{*}}{l}{S}{n}\cg{V^{*}}{k}{W}{j}{S^{*}}{n} g_{S}^{vw, vw}(1-z, 1-\bar{z})\,,
    \)}
\end{equation}
and again using Eq.~\eqref{eq:cg-complex-conjugate} we reconstruct the projector from \( V \otimes W^{*} \to S \)
\begin{equation}
    \braket{v_{i}(x_{1}) w_{j}(x_{2}) v_{k}(x_{3})^{*} w_{l}(x_{4})^{*}} = \mathbf{K}_{14,32}\sum_{S \in V \otimes W^{*}} (P^{V \otimes W^{*}}_{S})_{iklj} g_{S}^{vw, vw}(1-z, 1-\bar{z}).
\end{equation}
The crossing equation is then
\begin{equation}
    \mathbf{K}_{14,32}\sum_{S \in V \otimes W^{*}} (P^{V \otimes W^{*}}_{S})_{iklj} g_{S}^{vw, vw}(1-z, 1-\bar{z}) = \mathbf{K}_{12,34}\sum_{R \in V \otimes W} (P^{V \otimes W}_{R})_{ikjl} g_{R}^{vw, vw}(z, \bar{z})\,.
\end{equation}
We contract the above equation with \( (P^{V \otimes W}_{R})_{ijkl} \), then on the RHS
\begin{equation}
    (P^{V \otimes W}_{R})_{ikjl}^{*} (P^{V \otimes W}_{R'})_{ikjl} = \delta_{R R'} \tr[P^{V \otimes W}_{R}] = \delta_{R R'} d_{R}\,,
\end{equation}
while on the LHS we get
\begin{equation}
    N_{RS} = (P^{V \otimes W}_{R})_{ikjl}^{*} (P^{V \otimes W^{*}}_{S})_{iklj}\,.
\end{equation}

We observe that even though we used multiple times that the representations are unitary, \( N_{RS} \) has a basis-independent expression. Firstly,
\begin{equation}
    N_{RS} = (P^{V \otimes W}_{R})^{*}_{ikjl} ({P}^{V \otimes W^{*}}_{S})_{iljk} = (P^{V \otimes W}_{R})^{*}_{jlik} ({P}^{V \otimes W^{*}}_{S})_{iljk} = \tr\left[P^{V \otimes W}_{R} \, {CP}^{V \otimes W}_{S}\right]\,,
\end{equation}
where \( CP \) stands for ``crossed projector'', i.e.
\begin{equation}
    ({CP}^{V \otimes W}_{S})_{ikjl} = ({P}^{V \otimes W^{*}}_{S})_{iljk} = \frac{d_{S}}{|G|} \sum_{g}\chi_{S}(g)^{*} \, V(g)_{ij} W(g)_{lk}^{*}\,.
\end{equation}
Now recall that for unitary representations \(W(g)_{lk}^{*} = W(g^{-1})_{kl} = W^{-1}(g)_{kl}\), thus
\begin{equation}
    {CP}^{V \otimes W}_{S} = P^{V \otimes W^{-1}}_{S}\,.
\end{equation}
Therefore
\begin{equation}
    N_{RS} = \tr\left[P^{V \otimes W}_{R}  \, P^{V \otimes W^{-1}}_{S} \right].
\end{equation}
\( N_{RS} \) can then be written in terms of characters as follows
\begin{align}
  N_{RS} = \frac{d_{R} d_{S}}{|G|^{2}} \sum_{g_{1}, g_{2}\in G} \chi_{R}(g_{1})^{*} \, \chi_{S}(g_{2})^{*} \, \chi_{V}(g_{1} g_{2}) \chi_{W}(g_{1} g_{2}^{-1})\,.
\end{align}
The crossing equation can thus be written as
\begin{equation}\label{eq:sum-rules-group-theory}
    \mathbf{K}_{14,32} \sum_{S \in V \otimes W^{*}} M_{RS} g_{S}^{vw,vw}(1-z, 1-\bar{z}) - \mathbf{K}_{12,34} g_{R}^{vw,vw}(z, \bar{z}) = 0\,, \qquad \forall R \in V \otimes W\,,
\end{equation}
where
\begin{equation}
    M_{RS} = \frac{N_{RS}}{d_{R}} = \frac{d_{S}}{|G|^{2}} \sum_{g_{1}, g_{2}\in G} \chi_{R}(g_{1})^{*} \, \chi_{S}(g_{2})^{*} \, \chi_{V}(g_{1} g_{2}) \chi_{W}(g_{1} g_{2}^{-1})\,.
\end{equation}

We can take advantage of the fact that characters are constant over conjugacy classes to make the above formula more computationally approachable with GAP:
\begin{equation}
    M_{RS} = \frac{d_{S}}{|G|^{2}} \sum_{[i_{1}], [i_{2}] \in \mathrm{c.c.}} \chi_{R}([i_{1}])^{*} \, \chi_{S}([i_{2}])^{*} \left(\sum_{g_{1} \in i_{1}, g_{2} \in i_{2}} \chi_{V}(g_{1} g_{2}) \chi_{W}(g_{1} g_{2}^{-1})\right)\,,
\end{equation}
where \( i_{1}, i_{2} \) runs over the conjugacy classes of the group.

Computing the matrix $M_{RS}$ in this way, upon changing to the basis of convolved conformal blocks, we have checked the sum rules of Eq.~\ref{eq:sum-rules-group-theory} for the correlators $\braket{\phi\phi\phi\phi}$, $\braket{XXXX}$, $\braket{Z_3 Z_3 Z_3 Z_3}$, $\braket{\phi X\phi X}$, $\braket{\phi Z_3\phi Z_3}$ match the ones previously obtained by writing the explicit tensor structures.

{
\bibliographystyle{JHEP.bst}
\bibliography{Refs}
}

\end{document}